%% file: finfout.tex
\documentclass[11pt]{article}

\usepackage{booktabs} 

\pdfoutput=1
\usepackage{latexsym}
\usepackage{epsfig}
\usepackage[mathscr]{eucal}
\usepackage{amsfonts}
\usepackage{amscd}
\usepackage{cite}
\usepackage{array}
\usepackage{bbold}
\usepackage{amssymb}
\usepackage{colordvi}
\usepackage[centertags]{amsmath}
\usepackage{enumerate}
\usepackage{graphicx}
\usepackage{booktabs}
\usepackage{theorem}
\usepackage[footnotesize]{caption}
\usepackage{soul}
\usepackage{url}
\usepackage[hidelinks]{hyperref}
\usepackage{mcite}
\usepackage{slashed}
\usepackage[dvipsnames]{xcolor}
\usepackage{bbm}
\usepackage[utf8]{inputenc}

\allowdisplaybreaks{}

\numberwithin{equation}{section}

\newcommand{\be}{\begin{equation}}
\newcommand{\ee}{\end{equation}}
\newcommand{\beq}{\begin{eqnarray}}
\newcommand{\eeq}{\end{eqnarray}}

\newcommand{\mass}[1]{m_{#1}}
\newcommand{\RNum}[1]{\uppercase\expandafter{\romannumeral #1\relax}}

\newcommand{\ind}[1]{(i \in \{#1\})}
\newcommand{\si}[1]{\text}
\newcommand{\SI}[1]{\text}
\newcommand{\DS}{H_\text{DS}}
\newcommand{\HD}{H_\text{DD}}
\newcommand{\AD}{A_\text{D}}

\newcommand{\SM}{H_\text{SM}}

\usepackage{comment}

\begin{document}

\title{Freeze-in as a Complementary Process to Freeze-Out}

\date{\today}
\author{
Rodrigo Capucha$^{1\,}$\footnote{E-mail: \texttt{rscapucha@fc.ul.pt}},
Karim Elyaouti$^{2\,}$\footnote{E-mail:
	\texttt{karim.elyaouti@partner.kit.edu}},
Margarete M\"{u}hlleitner$^{2\,}$\footnote{E-mail:
	\texttt{margarete.muehlleitner@kit.edu}},\\
Johann Plotnikov$^{2\,}$\footnote{E-mail:  \texttt{johann.plotnikov@online.de}},
Rui Santos$^{1,3\,}$\footnote{E-mail:  \texttt{rasantos@fc.ul.pt}}
\\[9mm]
{\small\it
$^1$Centro de F\'{\i}sica Te\'{o}rica e Computacional,
    Faculdade de Ci\^{e}ncias,} \\
{\small \it    Universidade de Lisboa, Campo Grande, Edif\'{\i}cio C8
  1749-016 Lisboa, Portugal} \\[3mm]
{\small\it
$^2$Institute for Theoretical Physics, Karlsruhe Institute of Technology,} \\
{\small\it Wolfgang-Gaede-Str. 1, 76131 Karlsruhe, Germany.}\\[3mm]
{\small\it
$^3$ISEL -
 Instituto Superior de Engenharia de Lisboa,} \\
{\small \it   Instituto Polit\'ecnico de Lisboa
 1959-007 Lisboa, Portugal} \\[3mm]
}

\maketitle

\begin{abstract}
\noindent
There are many extensions of the Standard Model with a dark matter (DM) candidate obtained via the freeze-out mechanism. It can happen that after all experimental and theoretical constraints are taken into account, all parameter points have a relic density below the experimentally measured value. This means that the models solve only partially the DM problem, and at least one more candidate is needed. In this work we show that it is possible to further
extend the model with a DM candidate obtained via the freeze-in mechanism to be in agreement with the relic density experimental measurement.
Once the relic density problem is solved with this addition, new questions are raised. This new model with at least two DM candidates could have a freeze-out undetectable DM particle both in direct and indirect detection. This could happen if the freeze-out DM particle would have a very low density. Hence, a collider DM hint via excess in the missing energy with no correspondence in direct and indirect detection experiments, could signal the existence of a Feebly Interacting Massive Particle (FIMP). Conversely, if a DM particle is found and a particular model can explain all observables except the correct relic density, an extension with an extra FIMP would solve the problem. The freeze-in DM candidate, due to the small portal couplings, will not change the remaining phenomenology. 
\end{abstract}

\thispagestyle{empty}
\vfill
\newpage
\setcounter{page}{1}

\section{Introduction}
\hspace{\parindent} 

Ever since Dark Matter (DM) became an unavoidable ingredient of any sound extension of the Standard Model (SM)
the concept of a Higgs portal~\cite{Patt:2006fw} has been used to bridge the visible sector with the new dark sector.
Portal models assume that the two sectors only interact via the Higgs doublet field from the SM. The easiest way
to conceal one sector from the other is to impose some kind of symmetry, being the discrete $Z_2$ the most common together with the continuum $U(1)$ symmetry.
However,  any symmetry that stabilises the lightest neutral particle from the dark sector making it a viable DM candidate, is able to implement  the concept.

Assuming that DM is indeed a particle, there are two main mechanisms of DM production in agreement with all observations
and in particular with the observed relic density measured by PLANCK~\cite{Planck:2018vyg}.
In the freeze-out~\cite{Zeldovich:1965gev, Bertone:2004pz} mechanism DM is in thermal equilibrium with the thermal bath and when the temperature 
drops below the DM mass, the rate of expansion of the universe eventually becomes larger than the DM annihilation rate to lighter particles, and the annihilation stops. The surviving DM particles are known as Weakly Interacting Massive Particles (WIMPs).
In the freeze-in mechanism~\cite{Hall:2009bx} the assumption is that out of equilibrium DM particles are produced via decay
 and/or annihilation of other particles from the model. Again, when a certain temperature is reached
, production stops and the co-moving DM number
density becomes constant. Due to the very weak couplings involved, which is the cause of DM not being in thermal equilibrium, these are 
known as  Feebly Interacting Massive Particles (FIMPs). Besides the two main mechanisms there are
a number of other interesting proposals in the literature~\cite{DEramo:2010keq, Hochberg:2014dra, Kuflik:2015isi, DAgnolo:2015ujb, Pappadopulo:2016pkp, Garny:2017rxs, DAgnolo:2020mpt, Smirnov:2020zwf, Fitzpatrick:2020vba, Kramer:2020sbb,
Hryczuk:2021qtz, Bringmann:2021tjr} which are mainly derivations of the two above when considering specific regions of the parameter space.

Although over the last years WIMPs are the main paradigm explored both in the literature and in all kinds of experiments, FIMPs have gained
 a new life in recent years mainly because no DM particle was found in the WIMP mass range that goes roughly from GeV to TeV. 
 In order to test the WIMP paradigm to the limit we have to explore at least the simplest models in any new approach to understand if we did not miss something along the way. 
Once we pass the threshold of about half the Higgs mass, the most important constraints on dark matter are the ones coming from direct detection 
experiments\cite{LZ:2022lsv, XENON:2023cxc}\footnote{Indirect detection experiments also play a role. However, for most models or at least for 
the bulk of the parameter space direct detection dominates.} (DD) and the ones provided by Planck that establishes the value of the relic density. If there is only one portal coupling we can look at these constraints in very simple terms: a large portal coupling
is needed to be in agreement with relic density measurements and a small portal coupling is the way to avoid the direct detection bounds.
If freeze-out is the preferred mechanism of dark matter production, the portal coupling has to find some space between these two constraints and if that is not possible the model is excluded.
In the case of freeze-in, since the bounds of DD do not apply, the model is usually sound although not easy to exclude by any of the other experimental results. 

 The simplest extensions of the SM, like the real singlet extension~\cite{McDonald:1993ex},
 are usually enough to provide a DM candidate in agreement with all experimental data both as a WIMP and as a FIMP
 in a wide DM mass range. However, present DD constraints together with the need of abiding by the measured relic density
 already exclude the freeze-out hypothesis for a dark matter mass below $\approx 4$ TeV as we will show later.
 Whatever the scalar portal extension of the SM is, DD is only affected by a limited number out of the possible portal couplings in the model.  
 The introduction
 of new fields in the same dark matter sector, with the same dark quantum number, enables the opening of co-annihilation channels, alleviating therefore the tension  between relic density and direct detection.  
 Such is the case of the Inert Doublet Model (IDM)~\cite{Deshpande:1977rw} with an enlarged dark sector, consisting
 of two neutral and two charged particles. Because the DM candidate comes from an $SU(2)_L$ doublet, freeze-in is forbidden. As for freeze-out,
 the co-annihilation channels do indeed ease the tension allowing for a wider mass range. 
 As reported in~\cite{Arhrib:2013ela, Ilnicka:2015jba, Belyaev:2016lok, Kalinowski:2018ylg, Engeln:2020fld} the 
dark matter relic density cannot be saturated for DM masses between about 100 and 500 GeV. The addition 
of a real singlet with a different dark quantum number allows to fill in the relic density gap. 
This is the Full Dark Phase (FDP) scenario~\cite{Engeln:2020fld} of the 
 Next-to-Minimal 2-Higgs-Doublet Model (N2HDM)~\cite{Chen:2013jvg,Drozd:2014yla,Muhlleitner:2016mzt, Engeln:2020fld},
where both freeze-in and freeze-out can occur, leading to the measured value of the experimentally measured relic density in a complementary way. 

As it was briefly discussed in~\cite{Belanger:2021lwd, Belanger:2022qxt} multi-component dark matter models may have two
WIMP candidates but also one WIMP and one FIMP DM particle. In the latter scenario 
 the relic density can still be saturated while the WIMP DM candidate
respects the constraints from direct and indirect detection experiments. In~\cite{Belanger:2022qxt} this possibility
was explored with a $Z_4$ symmetry for the IDM plus a singlet. The same idea was discussed in~\cite{Bhattacharya:2021rwh} for a scenario with a vector and a scalar DM particle 
where the four possibilities of WIMP/FIMP were examined. 

In this work we examine the complementarity between freeze-out and freeze-in in models which can have two dark matter candidates. One of the particles is produced via freeze-out
and the other via freeze-in. Because FIMPs are hard to detect they could reveal themselves by the fact that they are instrumental in obtaining the correct relic density. 
Suppose that a dark matter particle is found in a DD experiment, and that LHC searches (or searches at future colliders) point to a given extension of the scalar sector of the SM,
allowing to determine couplings and masses of the model. If the relic density is not saturated this could hint to a FIMP being present in the model. Also, a DM hint at the LHC, having no correspondence in direct and indirect detection experiments in extensions of the SM with DM being produced via freeze-out could be solved with the addition of a FIMP candidate.
We will show how this complementarity works  in detail using a multi-component dark matter model with two real singlets with two independent $Z_2$ symmetries.
We will then move to more elaborated extensions of the SM. As a side remark we will also show the difference in having one or two DM candidates and
review some of the possible mechanisms proposed in the literature for the two scenarios.

The paper is organised as follows. In Section~\ref{sec:2} we discuss a toy model to make easier the understanding of the main points we want to make.
In Section~\ref{sec:3} we present the full dark phase of the N2HDM. Our conclusions are given in Section~\ref{sec:conclusions}.

\section{Two Real Singlets Extension of the SM}
\label{sec:2} 

The simplest extension of the SM that one can build to show the freeze-out and freeze-in complementarity is the addition of two singlets to the SM field content, that will play the role of DM candidates.
So let us consider a scalar potential where besides the SM Higgs doublet we have two real singlets and two discrete symmetries  $\mathbb{Z}_2^{\text{FO}}$
and $\mathbb{Z}_2^{\text{FI}}$  under which the potential is invariant.
Only  $\phi_{\text{FI}}$ is odd under  $\mathbb{Z}_2^{\text{FI}}$ and only  $\phi_{\text{FO}}$ is odd under  $\mathbb{Z}_2^{\text{FO}}$. The most general renormalisable scalar potential is then 
\begin{align}
V_{\text{Scalar}} = & \enspace \mu_{h}^{2} |H|^2 + \lambda_h |H|^4 + \frac{1}{2} \mu_{\text{FO}}^2 \, \phi_{\text{FO}}^2 +  \frac{\lambda_{1}}{4!} \, \phi_{\text{FO}}^4  + \frac{1}{2}\mu_{\text{FI}}^2 \, \phi_{\text{FI}}^2 +  \frac{\lambda_{2}}{4!} \, \phi_{\text{FI}}^4 
\nonumber
\\
+ &  \enspace \frac{\lambda_{\text{FO}}}{2} \, \phi_{\text{FO}}^2 |H|^2  + \frac{\lambda_{\text{FI}}}{2} \, \phi_{\text{FI}}^2 |H|^2 +  \frac{\lambda_{\text{IO}}}{4} \, \phi_{\text{FI}}^2 \phi_{\text{FO}}^2 \;, 
\end{align}
where $H$ is the SM Higgs doublet and  $\phi_{\text{FI}}$  and  $\phi_{\text{FO}}$ are the two real singlets. As the two symmetries remain unbroken, we end up with two DM candidates with masses
$m_{\text{FO}}$ and $m_{\text{FI}}$. The SM couplings are unchanged and the DM particles only couple to the Higgs boson via the Higgs portal, at tree-level. 

\begin{figure}[h!]
    \centering
    \includegraphics[width=0.65\linewidth]{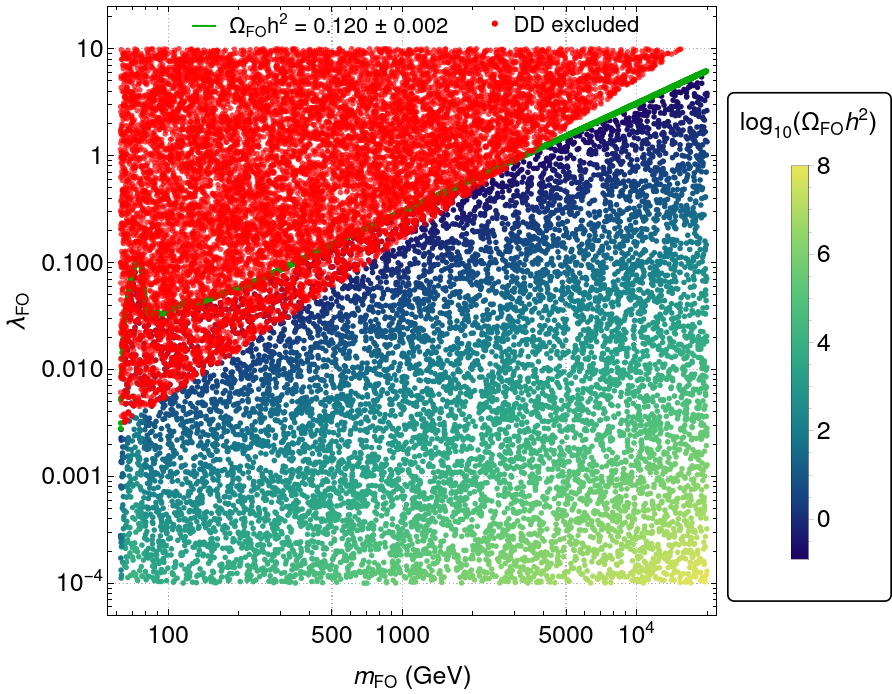}
    \caption{The white region shows the allowed parameter space in the
      plane ($\lambda_{\text{FO}}, \, m_{\text{FO}}$) in the scenario
      where $\lambda_{\text{FI}}=\lambda_{\text{IO}}=0$, for a dark
      matter mass for the particle that freezes out between $m_h/2$ and 20 TeV, and for $\lambda_{\text{FO}}$ between $10^{-4}$ and 10. The scenario corresponds to having only one singlet. The constraints shown are the strongest ones for this region of the parameter space: the latest direct detection measurements and the relic density. The red points are excluded by DD experiments, the dark green line describes points with the experimentally measured relic density, and all other points are above that value. The colored bar shows the freeze-out relic density for each point, $\Omega_{\text{FO}} h^2$, in logarithmic scale.}
    \label{fig:FO}
\end{figure}

The two limits  $\lambda_{\text{FI}}=0$ and  $\lambda_{\text{FO}}=0$ (with $\lambda_{\text{IO}} = 0$ in both cases)
correspond to scenarios where only one singlet is active and DM is generated via freeze-out and via freeze-in, respectively. Before discussing
the complementary we note that if  $\lambda_{\text{FO}}=0$, the correct DM density can be generated via freeze-in. In this case, since the coupling is extremely small, it will be difficult to test the model by any
other means. If however $\lambda_{\text{FI}}=0$ and the DM candidate
is generated via freeze-out the model is excluded in a vast region of
the parameter space. In Fig.~\ref{fig:FO} we show in white the
allowed region in the plane ($\lambda_{\text{FO}}, \, m_{\text{FO}}$)
in the scenario where $\lambda_{\text{FI}}=0$ and
$\lambda_{\text{IO}}=0$ (the remaining quartic couplings, $\lambda_1$
and $\lambda_2$, are irrelevant for the relic density, as well as
$\mu_{\text{FI}}$ in this case). The mass range chosen for the DM
particle that freezes out, $m_{\text{FO}}$, between $m_h/2$ and 20 TeV, is the one for which the
interplay between direct detection and relic density constraint is clearest.  The freeze-out portal coupling, $\lambda_{\text{FO}}$, varies between $10^{-4}$ and 10. As the coupling grows it becomes more and more constrained by the DD experiments. These experiments work optimally in the mass region
between 10 and 100 GeV and for a mass of 10 TeV the constraints are weaker by about two orders of magnitude. In order not to produce an overabundant relic density, one needs a large value of the portal coupling. The tension
between these two constraints leads to a small region of allowed couplings for masses above about 4 TeV. Future DD constraints will further reduce this region and make the searches at the LHC harder. In fact, for a mass of 4 TeV the DM production cross section is already negligible.
The plot was obtained using {\tt micrOMEGAs}~\cite{Belanger:2013oya, Belanger:2018mqt}.

%
%

\subsection{Determining the Relic Density via Freeze-in and Freeze-out}
In the previous section we have seen that a DM candidate produced via freeze-out is excluded for a mass below $\approx 4$ TeV in the single singlet extension of the SM with DM being produced via freeze-out. With the improved constraints on direct detection, the allowed mass range will soon be out of reach of the LHC. In this section we will discuss the complementarity between the two mechanisms. 
The coupled Boltzmann equations to determine the freeze-in and freeze-out yields for this scenario are
\begin{align}
 \frac{dY_\text{FI}}{dx} = & \sqrt{\frac{\pi}{45 G}} \, \frac{g_*^{1/2} m}{x^2}  \left[ \langle \sigma v \rangle_{\phi_{\text{FI}} \phi_{\text{FI}}  \text{SM}  \text{SM} } (Y_{\text{FI},\text{eq}}^2 - Y_\text{FI}^2)
\right.  \qquad  \qquad    \nonumber \\
& \left.  + \langle \sigma v \rangle_{\phi_{\text{FI}} \phi_{\text{FI}} \phi_{\text{FO}} \phi_{\text{FO}} } \left(  \frac{Y_\text{FO}^2}{Y_{\text{FO},\text{eq}}^2} Y_{\text{FI},\text{eq}}^2 -Y_\text{FI}^2
\right)
\right] \, ,
\end{align}
\begin{align}
\frac{dY_\text{FO}}{dx} = &  \sqrt{\frac{\pi}{45 G}} \, \frac{g_*^{1/2} m}{x^2}  \left[ \langle \sigma v \rangle_{\phi_{\text{FO}} \phi_{\text{FO}}  \text{SM}  \text{SM} } (Y_{\text{FO},\text{eq}}^2 - Y_\text{FO}^2) 
\right.  \qquad  \qquad    \nonumber \\
& \left.  - \langle \sigma v \rangle_{\phi_{\text{FI}} \phi_{\text{FI}} \phi_{\text{FO}} \phi_{\text{FO}} } \left(  \frac{Y_\text{FO}^2}{Y_{\text{FO},\text{eq}}^2} Y_{\text{FI},\text{eq}}^2 -Y_\text{FI}^2
\right)
\right] \, ,
\end{align}
where $G$ is the gravitational constant and $g_*$ are the relativistic degrees of freedom. The first equation determines the density of $\phi_{\text{FI}}$ via freeze-in. The quantity $\langle \sigma v \rangle_{\phi_{\text{FI}} \phi_{\text{FI}}  \text{SM}  \text{SM} }$ is the thermally averaged cross-section (TAC) for the $2 \to 2$ processes 
between $\phi_{\text{FI}}$ and the SM particles, where the sum over all possible SM final states is implicit (same definition for the $\text{FO}$ case).
The quantity $\langle \sigma v \rangle_{\phi_{\text{FI}} \phi_{\text{FI} } \phi_{\text{FO}} \phi_{\text{FO} }}$  is the TAC for the annihilation of two $\phi_{\text{FI}}$ into two DM particles odd
under $\mathbb{Z}_2^{\text{FO}}$. $Y_{\text{FO/FI},\text{eq}}$ are the
freeze-out/freeze-in equilibrium yields. In principle, we could also
add a decay term to the freeze-in equation. However, we do not include this term since we will work with masses for the particle that freezes in, $m_{\text{FI}}$, which make the decay kinematically forbidden. Note also that the $\langle \sigma v \rangle_{\phi_{\text{FI}} \phi_{\text{FI} } \phi_{\text{FO}} \phi_{\text{FO} }}$ TAC is multiplied by what we define as an enhancement factor, $\overline{Y}^2$, given by
\begin{equation}
    \overline{Y}^2 \equiv
    \frac{Y_{\text{FO}}^2}{Y_{{\text{FO}},\mathrm{eq}}^2} \;.\label{enhr}
\end{equation}
This enhancement factor is a measure of how much bigger the density of the particle that freezes out is compared to its equilibrium value. Because the SM is always in thermal equilibrium during freeze-out (and freeze-in) the corresponding terms do not have such enhancement factors.


%
The Boltzmann equations can be simplified by assuming that the density of  $\phi_{\text{FI} }$ is much lower than its equilibrium value. This is a good approximation since  $\phi_{\text{FI} }$ starts with no initial abundance 
and remains below the equilibrium density throughout the whole process. Hence, the terms proportional to $Y_\text{FI}$  can be neglected. 
%

Freeze-out and freeze-in occur at different times of the evolution of the universe. Typical freeze-out temperatures $T_{\mathrm{FO}}$, which do not result in an overabundance of the relic density, are in the order of $x_{\mathrm{FO}}=m_{\mathrm{FO}}/T_{\mathrm{FO}}\approx23-28$ \cite{Bender:2012gc},  while freeze-in typically ends at temperatures $T_{\mathrm{FI}}$ given by $x_{\mathrm{FI}}=m_{\mathrm{FI}}/T_{\mathrm{Fi}}\approx 2-5$ \cite{Hall:2009bx}. 
Suppose that $m_{\text{FO}} \approx m_{\text{FI}}$. This implies, that throughout the whole freeze-in of  $\phi_{\text{FI}}$,  $\phi_{\text{FO}}$ is in thermal equilibrium and therefore $\overline{Y}_\mathrm{}^2=1$.
If, however, the masses of $\phi_{\text{FO}}$ and $\phi_{\text{FI}}$  are far apart such that the temperatures of freeze-out and freeze-in overlap, the enhancement factor becomes non-unity during freeze-in. In this case ($T_\mathrm{FO} = T_\mathrm{FI}$) one obtains

\begin{equation}
    \frac{m_{\text{FO}}}{m_{\text{FI}}}\approx4.6-14 \;,
\end{equation}
for the typical freeze-out and freeze-in temperatures mentioned above. On the other hand, such high mass ratios lead to a small $\langle \sigma v \rangle_{\phi_{\text{FI}} \phi_{\text{FI}} \phi_{\text{FO}} \phi_{\text{FO}} }$ TAC making the enhancement negligible. As a consequence we can set the enhancement factors to one, independently of when freeze-out occurs. This means that the equations decouple and can be written as
\begin{align}
\frac{dY_\text{FI}}{dx} &= \sqrt{\frac{\pi}{45 G}} \, \frac{g_*^{1/2} m_{\text{FI}}}{x^2}  \left[ \langle \sigma v \rangle_{\phi_{\text{FI}} \phi_{\text{FI}}  \text{SM}  \text{SM} } + \langle \sigma v \rangle_{\phi_{\text{FI}} \phi_{\text{FI}} \phi_{\text{FO}} \phi_{\text{FO}} }
\right] \, Y_{\text{FI},\text{eq}}^2 \, ,\\
\frac{dY_\text{FO}}{dx} &= -\sqrt{\frac{\pi}{45 G}} \, \frac{g_*^{1/2} m_{\text{FI}}}{x^2}   \langle \sigma v \rangle_{\phi_{\text{FO}} \phi_{\text{FO}}  \text{SM}  \text{SM}} \left(Y_{\text{FO}} - Y_{\text{FO},\text{eq}}^2\right) ,
\end{align}
where we additionally neglect the $\langle \sigma v
\rangle_{\phi_{\text{FI}} \phi_{\text{FI}} \phi_{\text{FO}}
  \phi_{\text{FO}} }$ term in the equation for freeze-out because
$\langle \sigma v \rangle_{\phi_{\text{FI}} \phi_{\text{FI}}
  \phi_{\text{FO}} \phi_{\text{FO}} } \, \ll \, \langle \sigma v
\rangle_{\phi_{\text{FO}} \phi_{\text{FO}}  \text{SM}  \text{SM}}$ due
to the differences in the coupling strengths between freeze-out and
freeze-in. Solving these equations we can find the regions of the parameter space that lead to the correct  measured value of the relic density, that is
\begin{equation}
(\Omega h^2)_{\text{Planck}} = (\Omega_{\text{FI}} + \Omega_{\text{FO}}) h^2 = 0.120 \pm 0.002 \; ,\label{eq:plank_om}
\end{equation}
where $\Omega_{\text{FI}} h^2$ and $\Omega_{\text{FO}} h^2$ are the relic densities generated via freeze-in and freeze-out, respectively. 
\begin{figure}[h!]
  \centering
\includegraphics[width=0.49\linewidth]{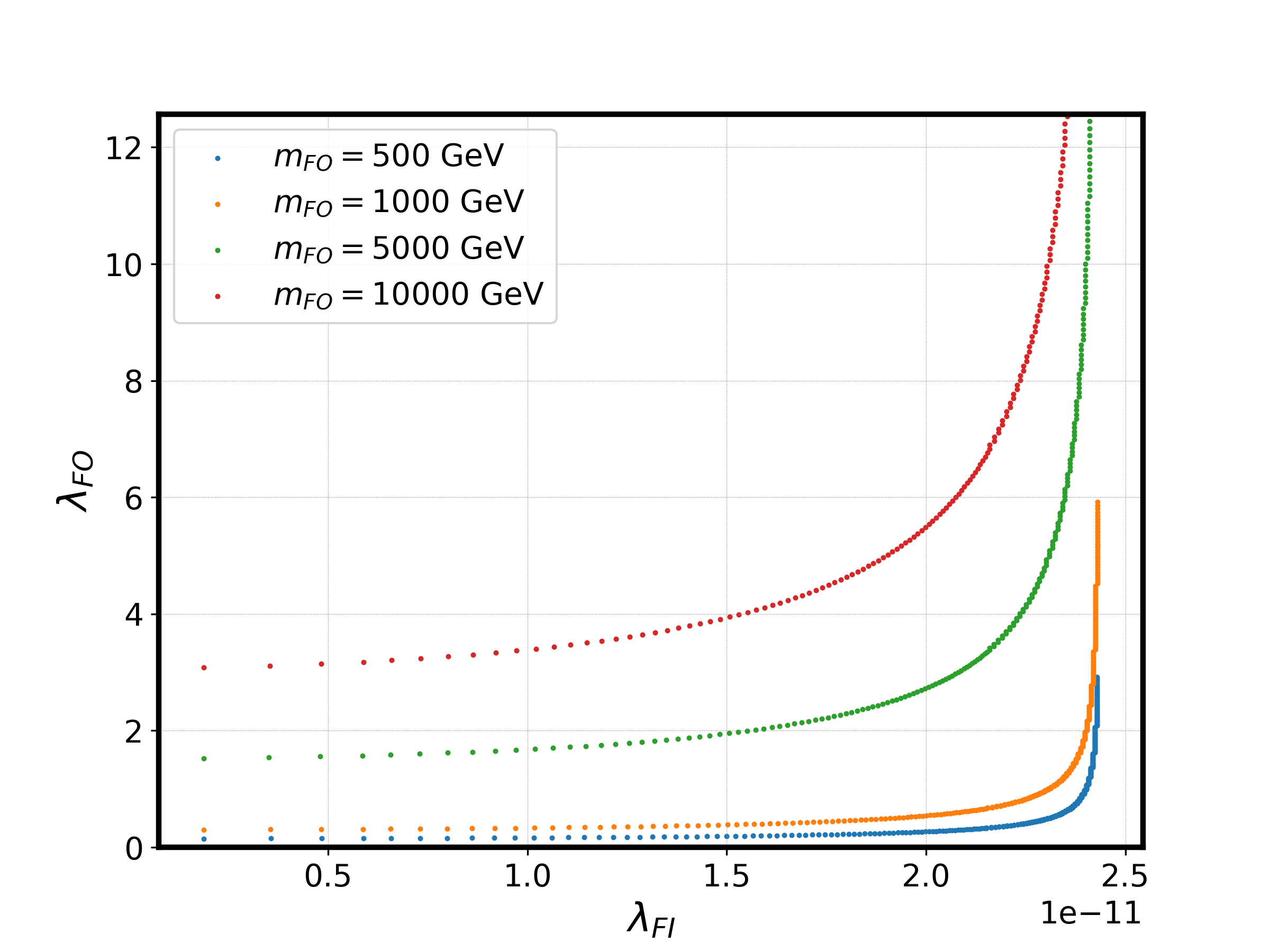}
\includegraphics[width=0.49\linewidth]{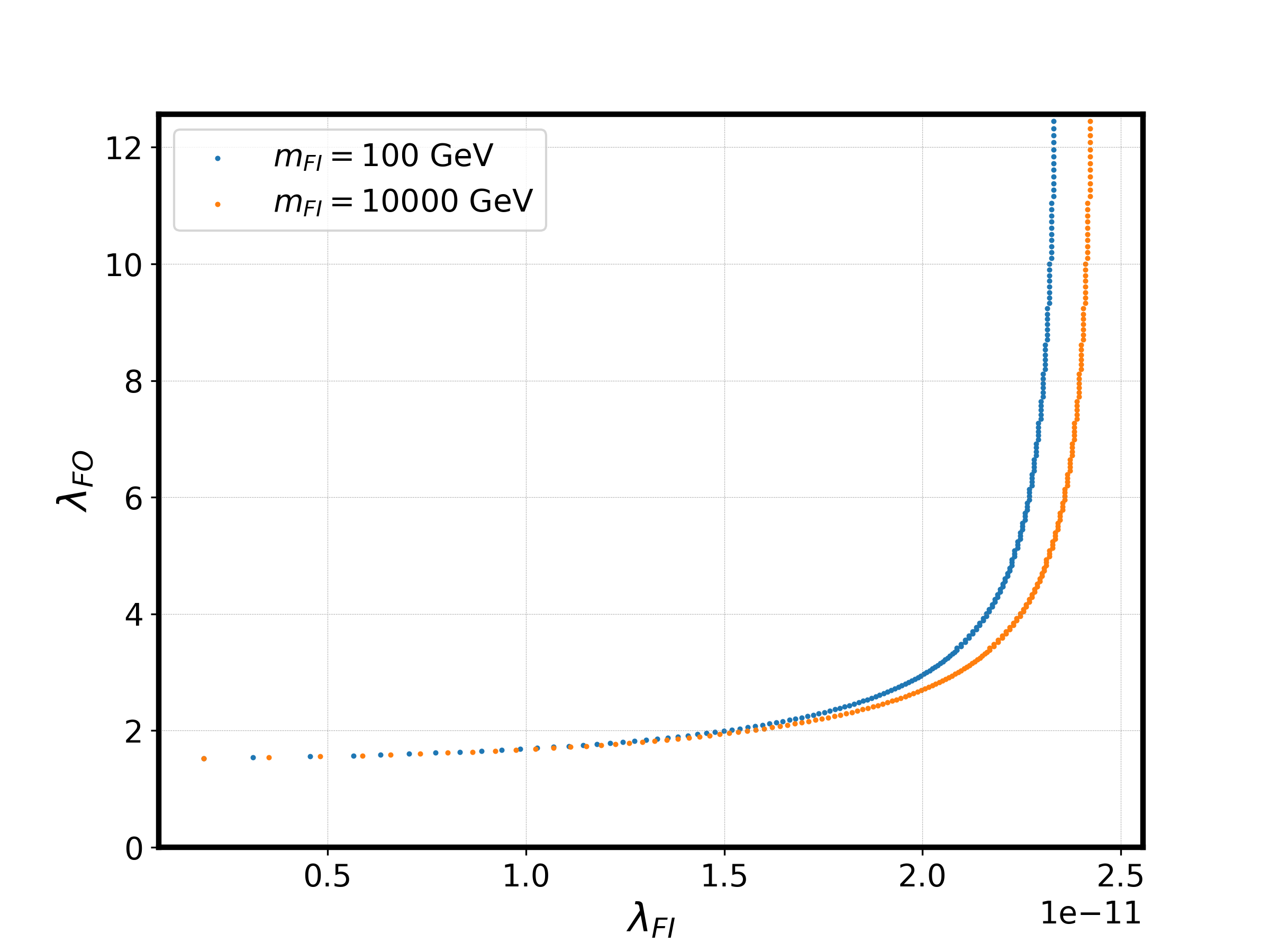}
  \caption{Lines for which $(\Omega h^2)_{\text{Planck}} =
    (\Omega_{\text{FI}} + \Omega_{\text{FO}}) h^2 = 0.120 \pm 0.002$
    in the ($\lambda_{\text{FO}}, \, \lambda_{\text{FI}}$) plane for
    several combinations of masses. $\lambda_{\text{IO}}=0$ and
      left: $m_{\text{FI}}=700$~GeV, right: $m_{\text{FO}}= 5$~TeV, and no other
    constraints were applied.   
    }\label{fig:FIvsFO}
\end{figure}

In Fig.~\ref{fig:FIvsFO} we present the lines for which the relic density is saturated, that is, $(\Omega h^2)_{\text{Planck}} = (\Omega_{\text{FI}} + \Omega_{\text{FO}}) h^2 = 0.120 \pm 0.002$, in the ($\lambda_{\text{FO}}, \, \lambda_{\text{FI}}$) plane 
for several combinations of masses and for $\lambda_{\text{IO}}=0$. In
the left plot the freeze-in mass is fixed to 700 GeV, and in the right
plot it is the freeze-out mass that is fixed to 5 TeV to avoid the DD
constraints.\footnote{Note that the points corresponding to
  $m_\text{FO} =500$ GeV  and $m_\text{FO} =1000$ GeV are excluded by
  DD constraints as can be seen in Fig.~\ref{fig:FO}.}
In the extreme left of the left plot  we recover the scenario where the freeze-in portal coupling is zero. In that limit we just see the relation between the freeze-out portal coupling and the mass, showing that as the mass grows the coupling
  also grows, as we can see in Fig.~\ref{fig:FO}. In fact, as
  discussed in~\cite{Hall:2009bx}, the freeze-out relic density is
  proportional to $m_{\text{FO}}^2/\lambda^2_{\text{FO}}$ and a larger
  mass requires a larger coupling. The freeze-in relic density is
  proportional to $\lambda^2_{\text{FI}}$ with a negligible mass
  dependence which is clear from the right plot. This means that
  larger freeze-out masses need larger couplings to
  stay below the measured relic density value of $0.120$, while the
  freeze-in relic density is independent of the mass up to threshold
  effects. Looking at Fig.~\ref{fig:FO}, we can see that when we only
  have freeze-out, for $m_\text{FO} = 5$ TeV, $\lambda_\text{FO}
  \approx 2$ to saturate the relic density. Thus, on the right plot of Fig.~\ref{fig:FIvsFO}, when the FI portal coupling is zero we should have $\lambda_\text{FO} \approx 2$. As we increase the contribution from FI to the total relic density, by increasing $\lambda_\text{FI}$, we need to decrease the contribution from FO, by increasing $\lambda_\text{FO}$. This explains why both $\lambda_\text{FO}$ and $\lambda_\text{FI}$ need to simultaneously increase (or decrease) so that we obtain the correct relic density.
%

\begin{figure}[h!]
  \centering
\includegraphics[width=0.65\linewidth]{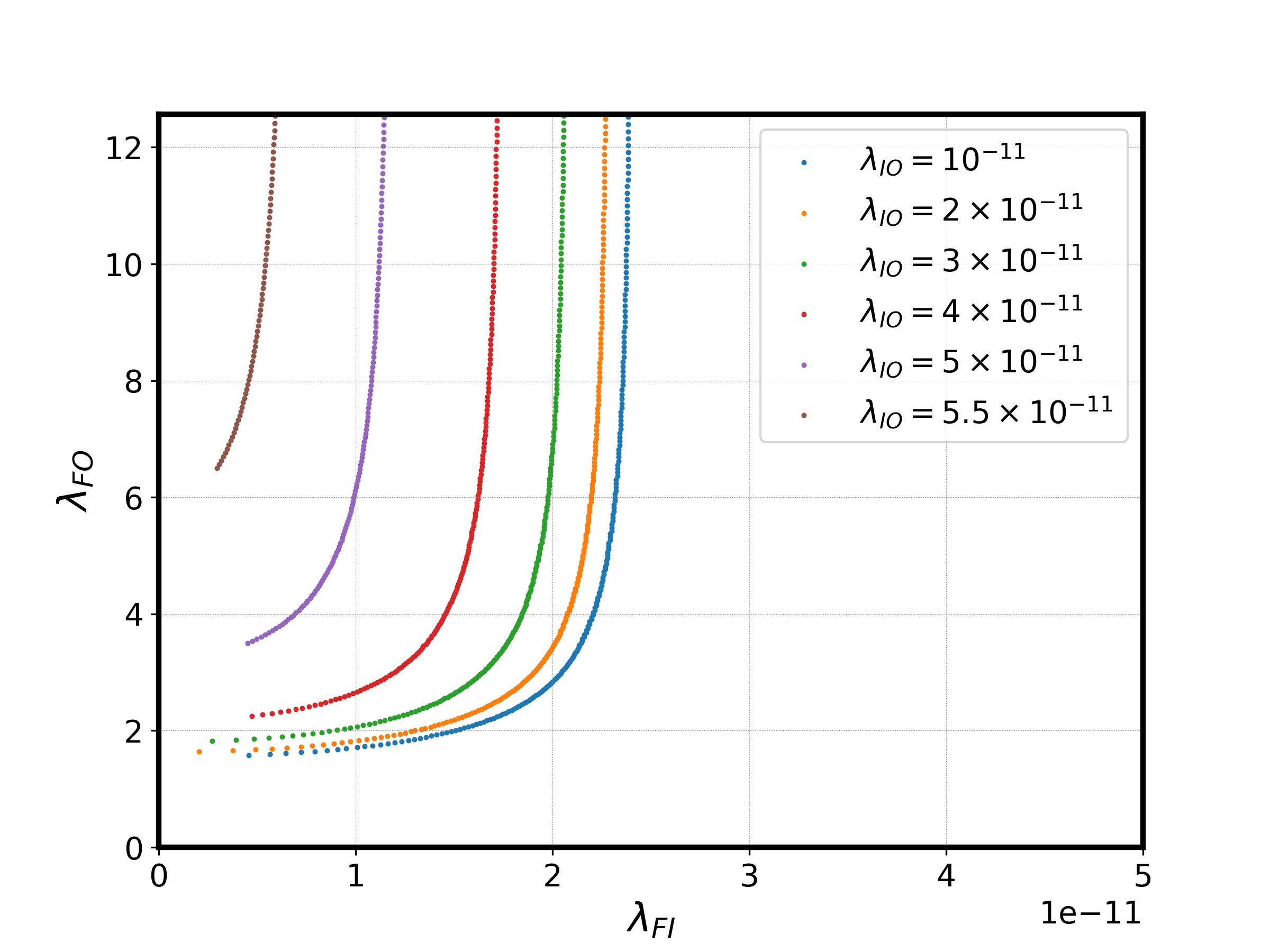}
  \caption{Lines for which $(\Omega h^2)_{\text{Planck}} = (\Omega_{\text{FI}} + \Omega_{\text{FO}}) h^2 = 0.120 \pm 0.002$ in the ($\lambda_{\text{FO}}, \, \lambda_{\text{FI}}$) plane for several values
  of $\lambda_{\text{IO}}$. No other constraints were applied
  apart from fixing the masses to $m_{\text{FO}} = 4$ TeV and $m_{\text{FI}} = 5$ TeV.
    }\label{fig:FOvsIO}
\end{figure}

In Fig.~\ref{fig:FOvsIO} we again present the lines
leading to $(\Omega h^2)_{\text{Planck}} = (\Omega_{\text{FI}} + \Omega_{\text{FO}}) h^2 = 0.120 \pm 0.002$ in the ($\lambda_{\text{FO}}, \, \lambda_{\text{FI}}$) plane.
In this case we fix the masses to be   $m_{\text{FO}} = 4$ TeV and
$m_{\text{FI}} = 5$ TeV and vary $\lambda_{\text{IO}}$.  As expected,
$\lambda_{\text{IO}}$ has to be of the order of $\lambda_{\text{FI}}$
because both couplings participate in the freeze-in process.
This plot clearly shows that by combining $\lambda_{\text{FI}}$ with
$\lambda_{\text{IO}}$ we can basically choose any value for the
freeze-out coupling, allowing in this way to consider parameter points
which were previously excluded when only FO was taken into
account. Again, this means that an invisible and undetected sector in
the model can always solve the issue of
obtaining the correct value of the relic density, provided it is underabundant. 
\\

\subsection{Consequences of Having Only One $\mathbb{Z}_2$ Symmetry}

\begin{figure}[h!]
    \centering
    
    \includegraphics[width=0.49\linewidth]{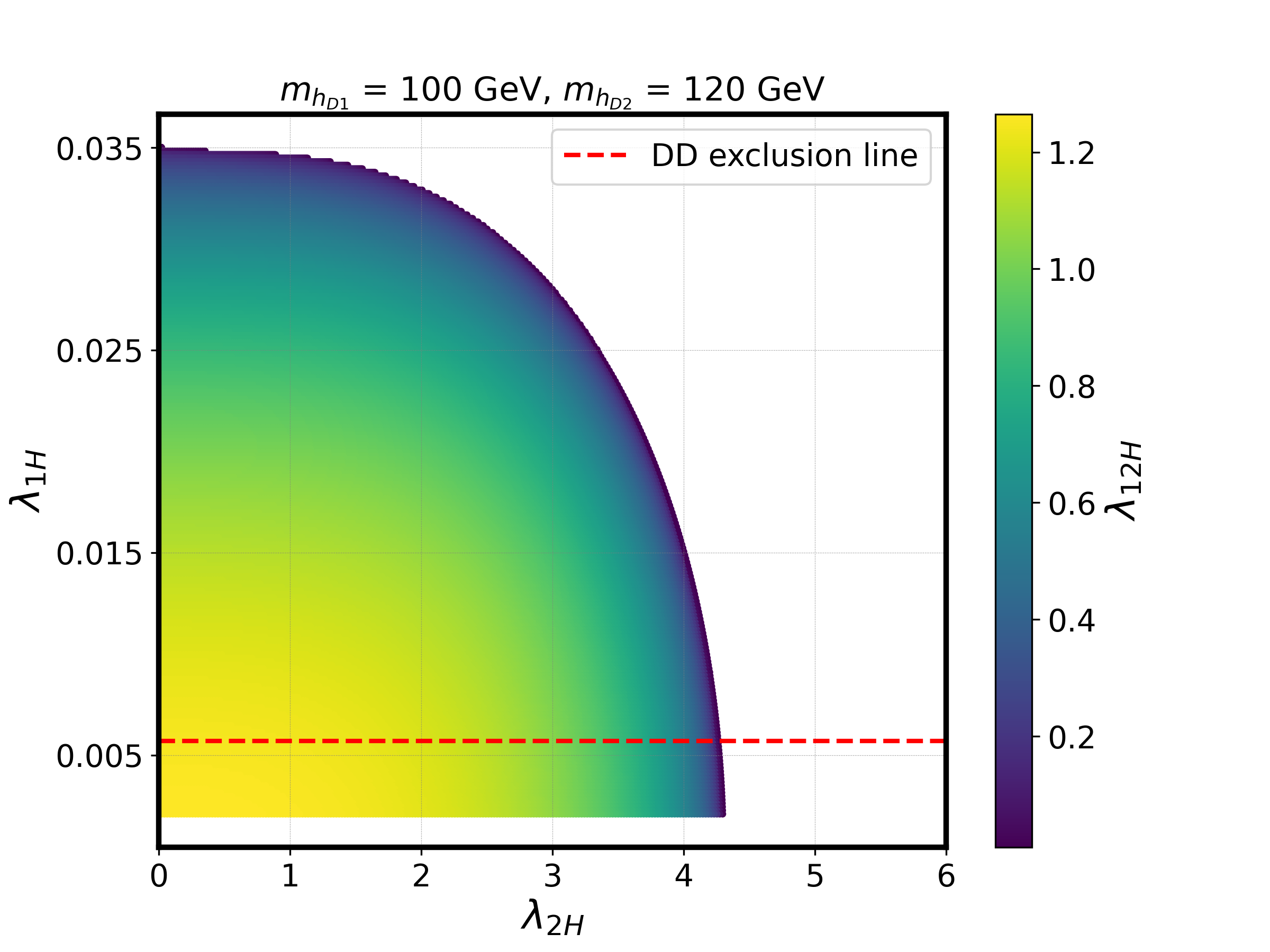}
    \includegraphics[width=0.49\linewidth]{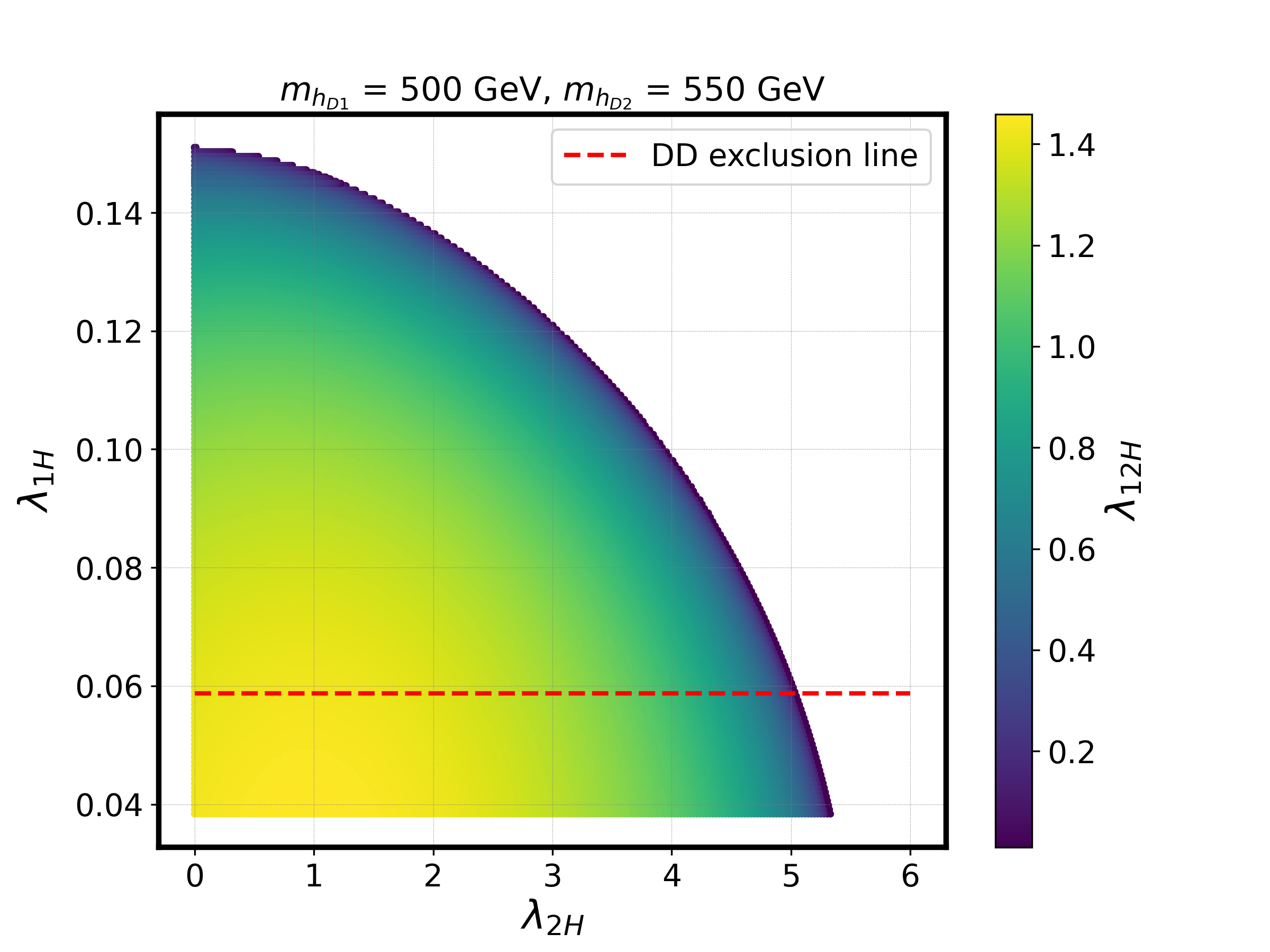}
    \includegraphics[width=0.49\linewidth]{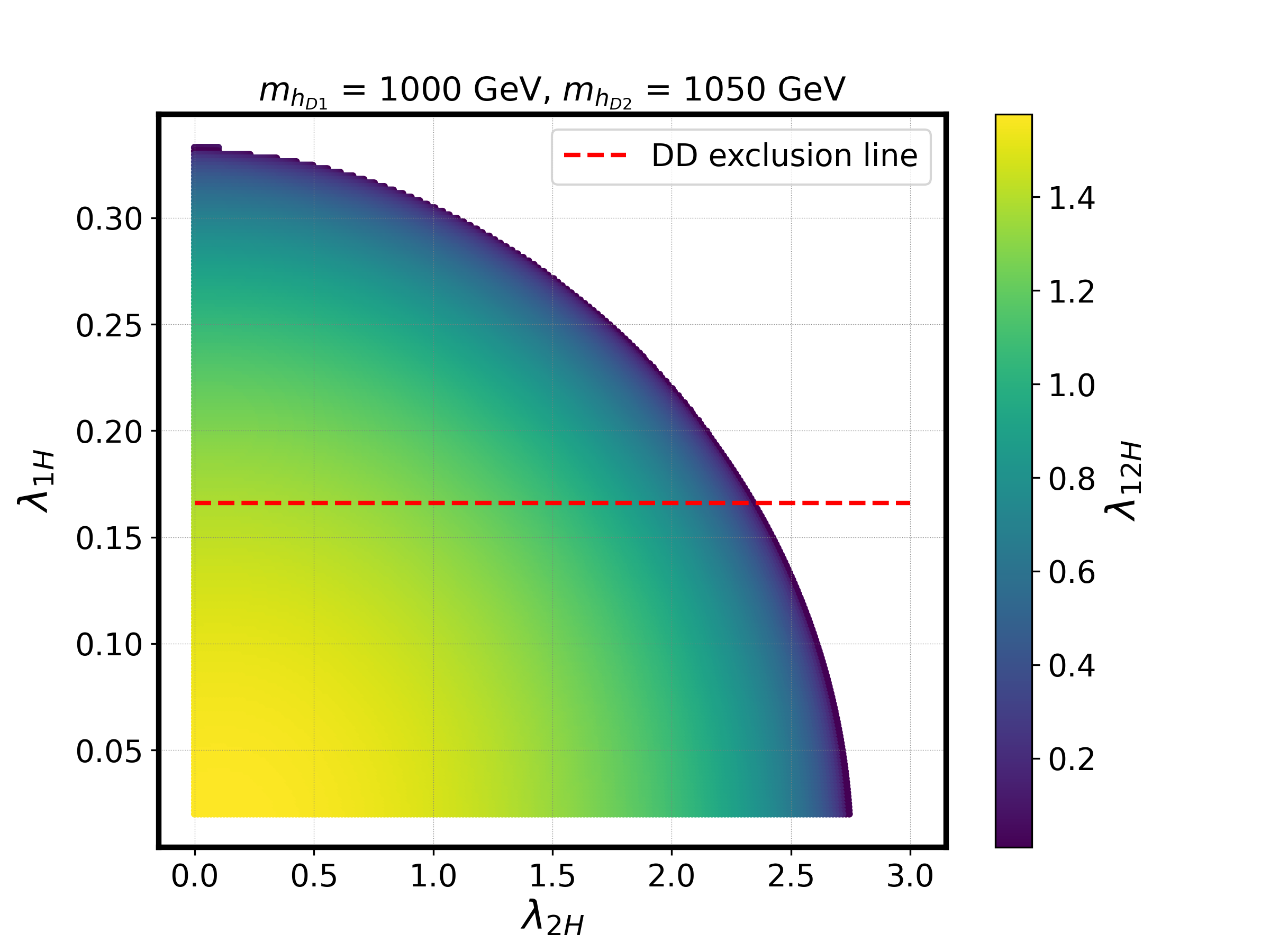}
    
    \caption{Parameter points for which $(\Omega h^2)_{\text{Planck}}
      = \Omega_\text{FO}h^2=0.120 \pm 0.002$ in the ($\lambda_{1H}$,
      $\lambda_{2H}$, $\lambda_{12H}$) plane for different
        $m_{h_{D1}}$, $m_{h_{D2}}$ mass combinations. The points above the red line are excluded by direct detection experiments. The remaining parameters of the potential are not relevant in this scenario.}
    
    \label{fig:FO_space}
\end{figure}

It is well-known that any extension of the SM depends heavily on the symmetries of its Lagrangian. Let us analyse the scenario where we again extend the model by two real singlets but now there is only one  $\mathbb{Z}_2$ symmetry. The most 
general renormalisable potential in this case is 
\begin{align}
V_{\text{Scalar}} = &\enspace \mu_{h}^{2} |H|^2 + \lambda_h |H|^4 +  m_1^2 \, \phi_1^2 +  \frac{\lambda_{1}}{4!} \, \phi_1^4  + m_2^2 \, \phi_2^2 +  \frac{\lambda_{2}}{4!} \, \phi_{2}^4 
\nonumber
\\
+ &  \enspace  \frac{\lambda_{1H}}{2} \, \phi_1^2 |H|^2  + \frac{\lambda_{2H}}{2} \, \phi_2^2 |H|^2 +  \frac{\lambda_{\text{12}}}{4} \, \phi_1^2 \phi_2^2 \label{eq:pot_1_Z2}\\
+ &  \enspace m_{12}^2 \phi_1 \phi_2  + \frac{\lambda_{\text{112}}}{4} \, \phi_1^3 \phi_2 +  \frac{\lambda_{\text{122}}}{4} \, \phi_1 \phi_2^3  +  \frac{\lambda_{12H}}{2} \, \phi_1 \phi_2  |H|^2 \, ,   \nonumber
\end{align}
where we have redefined the indices of the two scalar singlets so that they are now defined as $\phi_1$ and $\phi_2$. The last line of the potential shows the new terms that have to be added if only one discrete symmetry is present. 
Before we start the discussion we need to define the mass eigenstates in the dark sector (DS), which we call  $h_{D1}$ and $h_{D2}$. These are obtained from
$\phi_1$ and $\phi_2$ via the dark rotation angle $\theta_D$ as 
\begin{align}
\begin{pmatrix} h_{D1} \\ h_{D2}  \end{pmatrix} = 
\begin{pmatrix} \cos \theta_D & \sin \theta_D \\ -\sin \theta_D & \cos \theta_D
\end{pmatrix} \, \begin{pmatrix} \phi_1 \\ \phi_2  \end{pmatrix}.
\label{eq:phis}
\end{align}
In the case of having two $\mathbb{Z}_2$ symmetries as in the previous section it is easy to see how we are able to obtain 2 DM candidates which freeze-in and freeze-out respectively. However, the additional interactions introduced in this potential couple the two scalar fields, forcing them into thermal equilibrium with each other for typical freeze-out couplings. In this case we can describe the two scalar fields by one DS bath, and we will only have one DM candidate, corresponding to the lighter of the two DS particles. The Boltzmann equation describing the freeze-out of this DS bath is given by \cite{Edsjo:1997bg}
\begin{equation}
    \frac{dY}{dx}=-\sqrt{\frac{\pi}{45G}}\frac{g_*^{1/2} m_{h_{D1}}}{x^2}\langle\sigma v\rangle_\text{eff}\left(Y^2-Y_\text{eq}^2\right),
\end{equation}
where $m_{h_{D1}}$ is the mass of the lighter DS particle. In
$\langle\sigma v\rangle_\text{eff}$ we sum over all possible initial
and final states that contribute to freeze-out and in $Y_\text{eq}$ we
sum the equilibrium yields of both DS particles. By looking at the
potential in Eq.~(\ref{eq:pot_1_Z2}) we see that only the portal
couplings $\lambda_{1H}$, $\lambda_{2H}$ and $\lambda_{12H}$
contribute to freeze-out. This means
that we can draw a plane using these three parameters in which we
obtain the experimentally measured relic density given in
Eq.~(\ref{eq:plank_om}). This plane is shown in
Fig.~\ref{fig:FO_space}. Here we chose the mass of the DM candidate to
be $m_{h_{D1}} = 100$ GeV and the mass of the second scalar to be
$m_{h_{D2}}= 120$ GeV (upper left plot), $m_{h_{D1}} = 500$ GeV and
$m_{h_{D2}} = 550$ GeV (upper right plot), and $m_{h_{D1}} = 1000$ GeV
and $m_{h_{D2}} = 1050$ GeV (lower plot). The masses must be
relatively close to ensure that co-annihilations between the two DS
particles are relevant. The plane has three distinct points in which
each of the couplings is the dominant driver of the freeze-out process
such that $\frac{(\Omega h^2)_\text{total}}{(\Omega
  h^2)_\text{single}}>0.9$, where $(\Omega h^2)_\text{total}$ is the
relic density taking into account all processes, and $(\Omega
h^2)_\text{single}$ the relic density where we only consider the
processes associated to a specific single coupling. In the upper left, lower right and lower left corners of each plot in Fig.~\ref{fig:FO_space}, the dominant processes are $h_{D1}h_{D1}\rightarrow \text{SMSM}$, $h_{D2}h_{D2}\rightarrow \text{SMSM}$ and $h_{D1}h_{D2}\rightarrow \text{SMSM}$, respectively. This shows that without additional constraints there is a large and simple parameter space which allows to obtain the experimental relic density via FO alone, because now co-annihilation processes play a role, unlike in the previous scenario. Although we do not apply any other constraints it is easy to see how for example direct detection constraints would limit this parameter space. Since $\lambda_{1H}$ is the only relevant coupling for direct detection, there will be a clear upper bound on this coupling from such constraints.

Another way to obtain the full relic density in such a model would be again to let one of the DS particles freeze-in while the other one would freeze-out. This way we arrive at a similar situation as in the previous section in the case of degenerate masses. The only difference is that now there are more couplings that have to be small enough to enable freeze-in, which are $\lambda_{12}$, $\lambda_{112}$, $\lambda_{122}$, $\lambda_{12H}$ and $\lambda_{1H}$ or $\lambda_{2H}$. Additionally, there is also the possibility that the heavier of the DS particles can now decay into the lighter one via the coupling to the Higgs boson such that we only have one DM candidate. \\

\begin{figure}[h!]
  \centering
\includegraphics[width=0.6\linewidth]{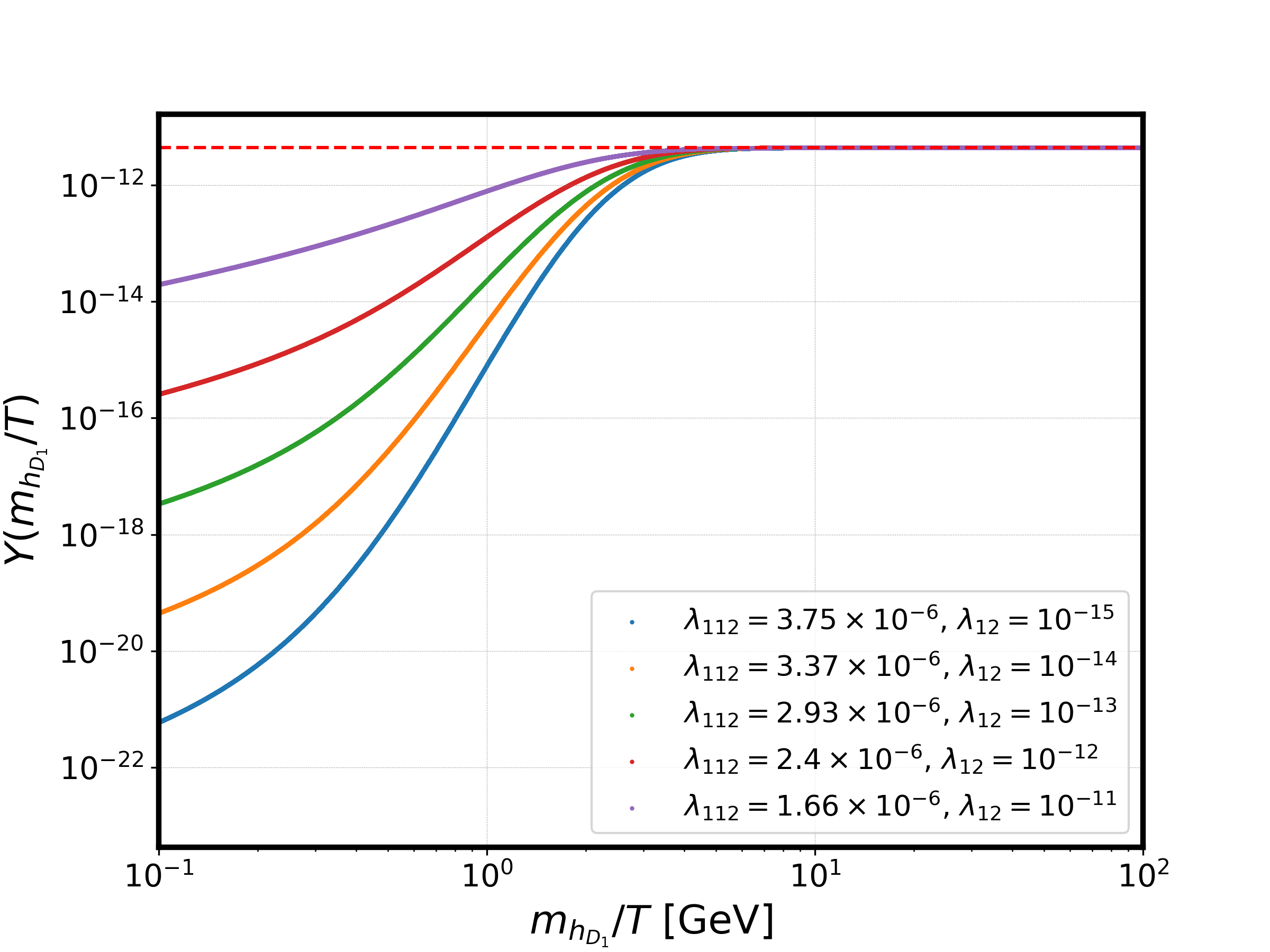}
  \caption{Yield of the DM particle $h_{D_1}$ as a function of
    $m_{h_{D_1}}/T$ for several combinations of the 
    potential quartic couplings $\lambda_{112}$ and
      $\lambda_{12}$. All other quartic couplings have been set
      to zero. The red dotted line corresponds to the yield for which we obtain the observed relic density.}
  \label{fig:expgrow}
\end{figure}

This model also allows for another mechanism to take place, the so
called DM from exponential growth, first introduced in
~\cite{Bringmann:2021tjr}. Although a specific model was not put
forward in~\cite{Bringmann:2021tjr} it can clearly be built with a
single $Z_2$ symmetry, like the model presented in this section. There
is also a study for a scenario with a $Z_3$
symmetry~\cite{Hryczuk:2021qtz}. In this production mechanism, at
least two particles from the DS are necessary. Of the two, the heavier
one (say $h_{D_2}$) is in thermal equilibrium with the SM bath and the
lighter one (say $h_{D_1}$), which is the DM candidate, starts with
zero initial abundance. The DM candidate then obtains a small
abundance through freeze-in via its interactions with the SM and the
heavier $h_{D_2}$ particle. With the right choice of couplings this
small abundance then grows exponentially through the process
$h_{D_2}h_{D_1}\rightarrow h_{D_1}h_{D_1}$ until it reaches a
plateau. This behaviour is shown in Fig.~\ref{fig:expgrow}, where the
lightest DM particle yield is shown as a function of $m_{h_{D_1}}/T$,
with $m_{h_{D_1}}=100$ GeV for several combinations of the
potential parameters that lead to the observed relic density. All of
the relevant quartic couplings except for $\lambda_{112}$ and
$\lambda_{12}$ were set to zero in this case. 
As expected, for a
smaller FI coupling, $\lambda_{12}$, the coupling associated with the
exponential growth, $\lambda_{112}$, needs to be larger, in order to
obtain the observed relic density. Also, the inclusion of the exponential growth mechanism allows to consider smaller FI couplings which are otherwise not able to reach the observed relic density.

In fact, as long as $\lambda_{12}$ is small enough so that FI alone
does not saturate the relic density, we can pratically choose any
value for the FI coupling and then adjust $\lambda_{112}$ accordingly
so that we have $\Omega h^2 = 0.120 \pm 0.002$. Another interesting
feature is the fact that, due to the exponential nature of this
mechanism, the coupling associated to exponential growth changes very
little. For instance, while $\lambda_{12}$ varies between four orders
of magnitude, $\lambda_{112}$ only changes by a factor of $\approx
2.26$ for the considered $\lambda_{12}$ values, meaning that this parameter is very constrained.

Assuming that $h_{D_2}$ remains in thermal equilibrium
 during the exponential growth of $h_{D_1}$ the evolution of the DM
 density for $h_{D_1}$ can be obtained from
\begin{equation}
\frac{\mathrm{d}Y_{h_{D_1}}}{\mathrm{d}x} = \sqrt{\frac{\pi}{45G}}
\frac{g_{*}^{1/2}m_{h_{D_2}}}{x^2}({\langle \sigma v
  \rangle_\text{exp} Y_{h_{D_2}}^{\text{eq}} Y_{h_{D_1}}} + \langle
\sigma v \rangle_\text{DS} (Y^\text{eq}_{h_{D_2}})^2+ \langle \sigma v
\rangle_\text{SM} (Y^\text{eq}_{h_{D_1}})^2) \;.\label{eq:exp}
\end{equation}
Here $Y_{h_{D_1}}^\text{eq}$ ($Y_{h_{D_2}}^\text{eq}$) is the
equilibrium yield of the particle $h_{D_1}$ ($h_{D_2}$), and the TACs
$\langle \sigma v \rangle_\text{exp}$, $\langle \sigma v
\rangle_\text{DS}$ and $\langle \sigma v \rangle_\text{SM}$ contribute
to exponential growth, freeze-in via the DS particle $h_{D_2}$
and freeze-in via the SM particles, respectively.

\begin{figure}[h!]
    \centering
    \includegraphics[width=0.49\linewidth]{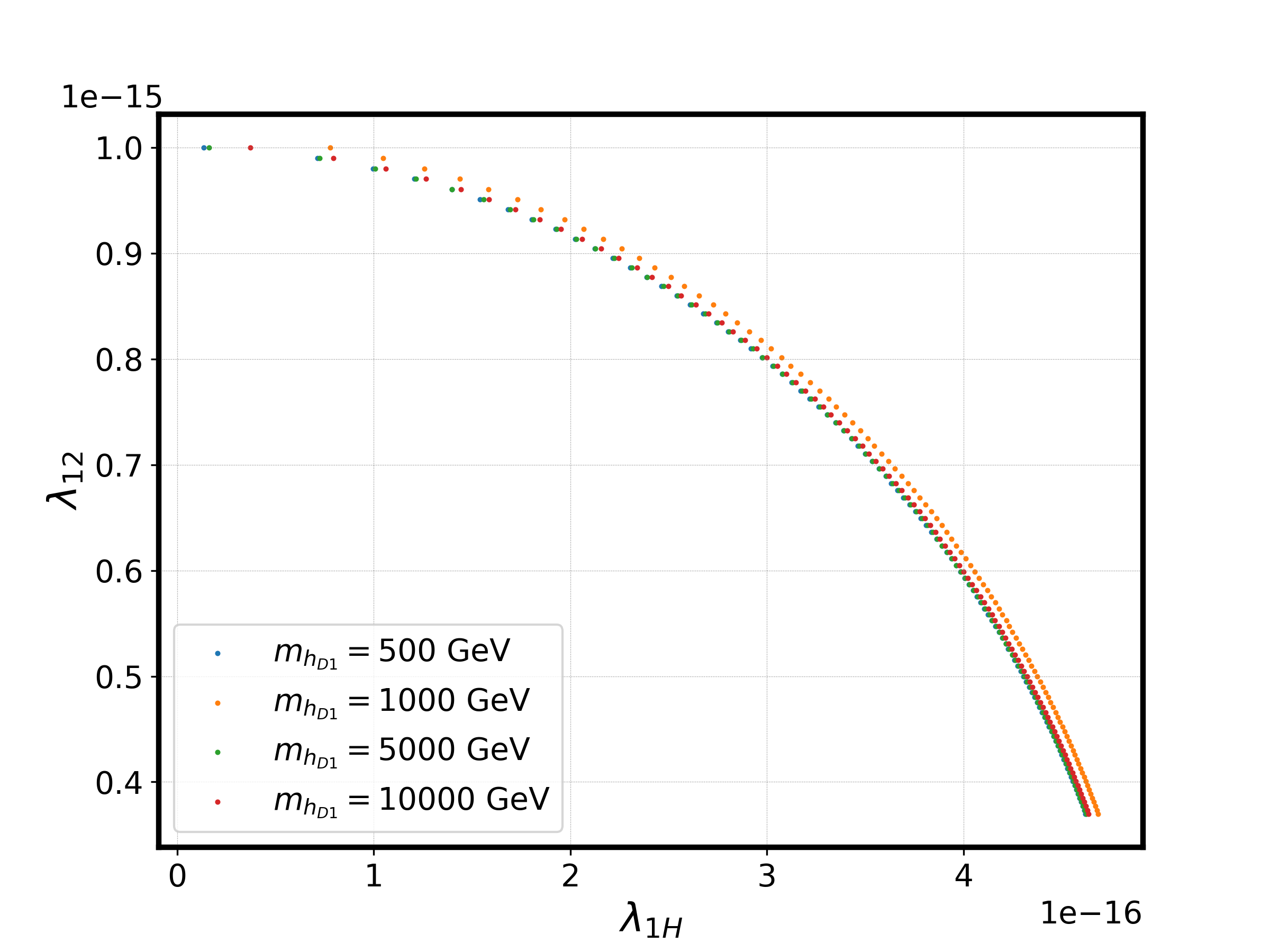}
    \includegraphics[width=0.49\linewidth]{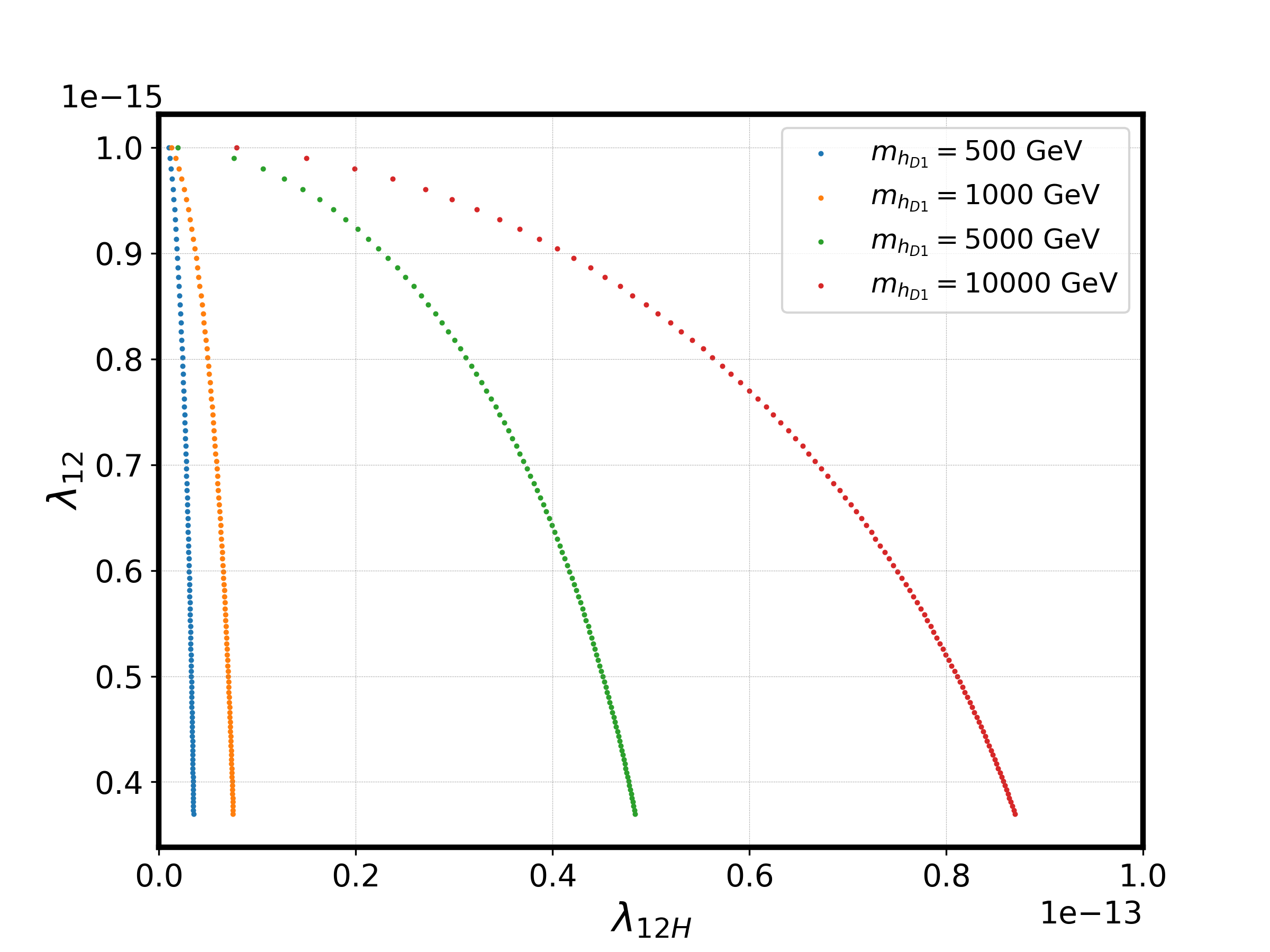}
    \includegraphics[width=0.49\linewidth]{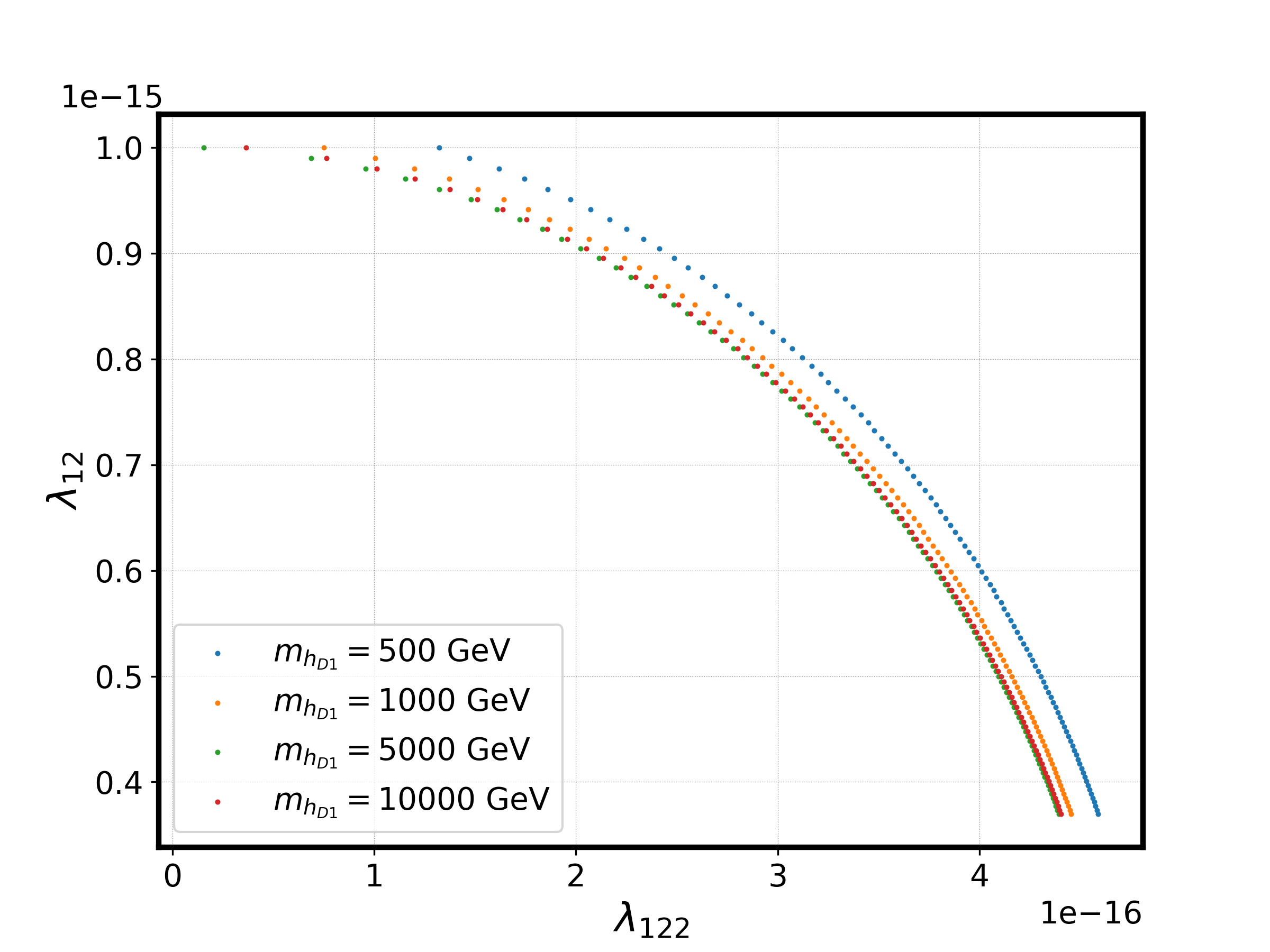}
    \caption{Parameter points for which the experimental relic density
      is obtained. The coupling $\lambda_{112}$ which is responsible
      for the exponential growth is fixed for each mass and
      $m_{h_{D_2}} = 2 \, m_{h_{D_1}}$.}
    \label{fig:exp_space}
\end{figure}

Again, it is easy to obtain a parameter combination
which is able to generate a relic density of $(\Omega
h^2)_\text{Planck}=0.120 \pm 0.002$ as can be seen in
Fig.~\ref{fig:exp_space}. Here, we fixed the coupling responsible for
exponential growth, $\lambda_{112}$, for each mass, such that for
$\lambda_{1H}=\lambda_{12H}=\lambda_{122}=0$ and
$\lambda_{12}=10^{-15}$ we obtain the observed relic density. We
  furthermore fix the DS masses such that $m_{h_{D_2}}= 2 m_{h_{D_1}}$. From
here we decrease the coupling $\lambda_{12}$ and adjust the other
freeze-in couplings and $\lambda_{112}$ such that we
still obtain the observed relic density. (The $\lambda_{112}$ coupling
changes by a very small amount due to the fact that it is related to
exponential growth, as we mentioned earlier.) As stated in the
previous section, these freeze-in couplings are too small to be
affected by other constraints. 

To summarize Sec.~\ref{sec:2}, we can see that by considering
different symmetries for 
a given Lagrangian and by moving into different parameter space
regions, we are able to obtain different DM generation scenarios. Here
we only focused on scenarios involving freeze-out or freeze-in with or without exponential growth. However, mechanisms such as Cannibal DM \cite{Pappadopulo:2016pkp} and Forbidden DM \cite{DAgnolo:2015ujb} would also allow to generate the experimentally measured relic density within this model, as well as semi-annihilation~\cite{DEramo:2010keq} in specific regions of the parameter space.

%

\section{The Full Dark Phase of the N2HDM}
\label{sec:3}

In this section we discuss a particular phase of the N2HDM~\cite{Chen:2013jvg,Drozd:2014yla,Muhlleitner:2016mzt, Engeln:2020fld}, where both FI and FO occur, and together reproduce the correct value of the experimentally measured relic density in a complementary way. The 
N2HDM has the SM particle content plus a second
$SU(2)_{L}$ doublet $\Phi_2$ with hypercharge $Y=+1$ and a real $SU(2)_{L}$ singlet $\Phi_S$ with $Y=0$.
The most general renormalisable scalar potential invariant under the two $\mathbb{Z}_{2}$ symmetries
\begin{eqnarray}
& & \mathbb{Z}^{(1)}_{2}:\quad \Phi_{1}\rightarrow \Phi_{1},\quad \Phi_{2}\rightarrow -\Phi_{2},\quad \Phi_{S}\rightarrow \Phi_{S}\,,\label{eq:Z2_1}\\
& & \mathbb{Z}^{(2)}_{2}:\quad \Phi_{1}\rightarrow \Phi_{1},\quad \Phi_{2}\rightarrow \Phi_{2},\quad \Phi_{S}\rightarrow -\Phi_{S}\label{eq:Z2_2} \, ,
\end{eqnarray}
is 
\begin{align}
V_{\text{Scalar}} =&\enspace m_{11}^{2} \Phi_{1}^{\dagger} \Phi_{1} + m_{22}^{2} \Phi_{2}^{\dagger} \Phi_{2}
+ \dfrac{\lambda_{1}}{2} \left(\Phi_{1}^{\dagger} \Phi_{1}\right)^{2}
+ \dfrac{\lambda_{2}}{2} \left(\Phi_{2}^{\dagger} \Phi_{2}\right)^{2}\notag\\
+&\enspace \lambda_{3} \Phi_{1}^{\dagger} \Phi_{1} \Phi_{2}^{\dagger} \Phi_{2}
+ \lambda_{4} \Phi_{1}^{\dagger} \Phi_{2} \Phi_{2}^{\dagger} \Phi_{1}
+ \dfrac{\lambda_{5}}{2} \left[\left(\Phi_{1}^{\dagger} \Phi_{2}\right)^{2} + \text{h.c.}\right]\label{eq:scalpot}\\
+&\enspace \dfrac{1}{2} m_{s}^{2}\Phi_{S}^{2} + \dfrac{\lambda_{6}}{8} \Phi_{S}^{4} + \dfrac{\lambda_{7}}{2} \Phi_{1}^{\dagger} \Phi_{1} \Phi_{S}^{2} + \dfrac{\lambda_{8}}{2} \Phi_{2}^{\dagger} \Phi_{2} \Phi_{S}^{2}\,,\notag
\end{align}
where all 11 free parameters of the Lagrangian,
\begin{equation}
  m_{11}^2\,,\enspace
  m_{22}^2\,,\enspace
  m_S^2\,,\enspace
  \lambda_{1-8}\,,\enspace\label{eq:lagrangepars}
\end{equation}
are real, or can be made to be so via a trivial rephasing of one of the doublets. We introduce no soft breaking terms
in the potential for the discrete symmetries  to be exact. Soft breaking mass terms are usually introduced in these models to allow for a decoupling limit. One such term is $m_{12}^2 ( \Phi_{1}^\dagger \Phi_2 + h.c.)$ usually present  in 2HDMs and N2HDMs.
Since this term would softly break  the $Z^{(1)}_2$ symmetry we will
not include it. Note that even if this particular term was included
the minimum condition would disallow it. 

As the two symmetries are exact both explicitly and spontaneously, two DM candidates will emerge after electroweak symmetry breaking (EWSB). The two DM quantum numbers
are independent and the corresponding DM candidates cannot decay into each other. This particular phase of the N2HDM was named in~\cite{Engeln:2020fld} the Full Dark Phase (FDP) of the N2HDM. A discussion of the different phases of the N2HDM can be found in~\cite{Engeln:2020fld}. 

After EWSB, the fields can be parametrised in terms of the charged complex fields $\phi_i^+$ $(i\in\{1,2\})$, the neutral CP-even fields $\rho_I$ $(I\in\{1,2,s\})$ and the neutral CP-odd fields $\eta_i$ as follows
\begin{align}
\Phi_1 = \begin{pmatrix} \phi_{1}^{+} \\
\dfrac{1}{\sqrt{2}}\left(v_{1} + \rho_{1} + i\,\eta_{1}\right)
\end{pmatrix},\quad
\Phi_2 = \begin{pmatrix} \phi_{2}^{+} \\
\dfrac{1}{\sqrt{2}}\left( \rho_{2} + i\,\eta_{2}\right)
\end{pmatrix}, \quad
\Phi_S =  \rho_{s}\,.
\label{eq:phis}
\end{align}
Because all other neutral fields belong to
one of the dark phases, the SM-like Higgs is the one from the doublet with
a vacuum expectation value (VEV). There is no mixing in the scalar sector and therefore the rotation matrix from the gauge eigenstates 
to the mass eigenstates is just the unit matrix, $\mathcal{R}=\mathbb{1}_{3\times3}$.
Also, the SM Higgs couplings to the other SM particles do not change relative to the SM. 
There are, however, new couplings, the ones between the Higgs and the dark matter candidates stemming
from the Higgs potential and the couplings of the dark particles from the doublet to the $SU(2)$ gauge bosons, via the covariant derivative. 
The mass eigenstates are the SM Higgs, $H_{\text{SM}}$, the four scalars from the dark doublet, $H_\text{DD}$, $A_\text{D}$, and the two charged scalars $H_\text{D}^\pm$, and the one from the dark singlet, $H_\text{DS}$.
We use as input values the following parameter set,
\begin{equation} \label{FDPinput}
    \mass{H_\text{SM}},\quad \mass{H_\text{DD}},\quad \mass{A_\text{D}},\quad \mass{H_\text{{DS}}},\quad \mass{H^\pm_\text{D}},\quad \mass{22}^2, \quad\lambda_2, \quad \lambda_6, \quad \lambda_7, \quad \lambda_8, \quad v .
\end{equation}
The choice of $\lambda_7$ instead of $m^2_\text{S}$ is motivated by
the fact that we can access the FI region via the parameter
$\lambda_7$ more directly than with $m^2_{S}$. They are related by the expression
\begin{equation}
    m^2_S = m_{H_\text{DS}} - \frac{\lambda_7}{2}v^2 \; .
  \end{equation}

Since the dark doublet particles couple to the $SU(2)$ gauge bosons
proportional to the $SU(2)_L$ and $U(1)_Y$ coupling constants $g$
and $g'$, respectively, these interactions have a fixed value. This in turn means
that neither $H_\text{DD}$ nor $A_\text{D}$ can be  
produced via the freeze-in mechanism, as these particles are in thermal equilibrium.
On the other hand, $H_{\text{DS}}$ can only couple to the SM via the SM Higgs boson as mediator. The trilinear coupling $\lambda(H_\text{DS},H_\text{DS},H_\text{SM})$ is given by
\begin{equation}\label{higgsportal}
    \lambda(H_\text{DS},H_\text{DS},H_\text{SM}) = -\lambda_7v \; .
\end{equation}
Therefore, the condition for freeze-in can only be accomplished if $\lambda_7$ is chosen to be small. Furthermore, the couplings of the doublet dark sector particles to $H_\text{DS}$ also have to be small, as these particles are in thermal equilibrium. 
A large coupling would also bring $H_{\text{DS}}$ to thermal equilibrium, and would therefore preclude freeze-in in the dark singlet sector. As the quartic coupling $\lambda(X,X,H_\text{DS},H_{\text{DS}})$ $(X \in \{H_{\text{DD}}, A_\text{D}, H_D^\pm\})$ is given by
\begin{equation}
    \lambda(X,X,H_\text{DS},H_{\text{DS}}) = - \lambda_8 \; ,
\end{equation}
we consequently choose $\lambda_8$ to be small as well. 

The dark sector from the doublet, similar to the one from the Inert
doublet model~\cite{Deshpande:1977rw, Ma:2006km, Barbieri:2006dq,
  LopezHonorez:2006gr}, will be responsible for freeze-out with
coupling constants of the order of $\lambda_i \sim 10^{-2} - 10^{-3}$
($i = 3-5$), and 
the dark sector from the singlet will have much smaller couplings so that freeze-in can occur.  This is the setup required to have simultaneously freeze-in and freeze-out.

\subsection{Constraints on the Model}
We will now describe briefly the theoretical and experimental constraints on the model and justify the choice for the regions of the parameter space scans. 
The constraints are imposed via the implementation of the model in $\texttt{ScannerS}$ \cite{Coimbra:2013qq, Muhlleitner:2020wwk}.

The necessary and sufficient conditions for the potential to be bounded from below were obtained in~\cite{Klimenko:1984qx} for a general N2HDM.
The discriminant $D$ defined in~\cite{Muhlleitner:2016mzt}, 
\begin{equation}
    D = \text{min}(\lambda_4-|\lambda_5|,0) \; ,
\end{equation}
allows to write the region of parameters for a potential bounded from below as
\begin{equation}
    \Omega_1 \cap \Omega_2\; ,
\end{equation}
where $\Omega_1$ and $\Omega_2$ are given by
\begin{equation}
\begin{split}
    \Omega_1 =& \Bigg\{ \lambda_1,\lambda_2,\lambda_6 > 0; \sqrt{\lambda_1\lambda_6}+\lambda_7 > 0;  \sqrt{\lambda_2\lambda_6}+\lambda_8 > 0;\\
    &\sqrt{\lambda_1\lambda_2} + \lambda_3 + D > 0; \sqrt{\frac{\lambda_1}{\lambda_2}}\lambda_8 + \lambda_7 \geq 0 \Bigg \} \; ,
    \end{split}
\end{equation}
\begin{equation}
\hspace{-3mm}
 \begin{split}
    \Omega_2 =& \Bigg\{ \lambda_1,\lambda_2,\lambda_6 > 0; \sqrt{\lambda_1\lambda_6}>-\lambda_7 \geq \sqrt{\frac{\lambda_1}{\lambda_2}}\lambda_8; \sqrt{\lambda_2\lambda_6}\geq\lambda_8>-\sqrt{\lambda_2\lambda_6}; \\
    &\sqrt{(\lambda_7^2 - \lambda_1\lambda_6)(\lambda_8^2 -\lambda_2\lambda_6)}>\lambda_7\lambda_8-(D+\lambda_3)\lambda_6 \Bigg \} \;.
 \end{split}
\end{equation}
To ensure at least (meta-)stability of the electroweak vacuum, we test each parameter point with $\texttt{EVADE}$~\cite{Hollik:2018wrr, Ferreira:2019iqb, Evade3}.

We require perturbative unitarity \cite{Lee:1977eg} to hold, meaning that the eigenvalues $\mathcal{E}_i$ of the tree-level $2\rightarrow2$ scattering matrix have to fulfil the following condition,
\begin{equation}
    |\mathcal{E}_i| \leq 8\pi \;.
\end{equation}
The eigenvalues $\mathcal{E}_i$ are given in Ref. \cite{Muhlleitner:2016mzt}. 

Adding additional scalars to the theory requires considering possible deviations from the electroweak precision parameters $S$, $T$, and $U$ \cite{Peskin:1991sw}. Compatibility with electroweak precision data is imposed by a 95\% confidence level (C.L.) exclusion limit using the formulae in Refs. \cite{Grimus:2007if, Grimus:2008nb} and the fit result of Ref. \cite{Haller:2018nnx}.
The formulae and constraints are implemented in  \texttt{ScannerS}.


The experiments at the Tevatron, LHC, and LEP provide additional
constraints on the new scalars of the model. These constraints are
tested via $\texttt{HiggsSignals}$ \cite{Bechtle:2013xfa, Bechtle:2014ewa}
and $\texttt{HiggsBounds}$
\cite{Bechtle:2008jh, Bechtle:2011sb, Bechtle:2012lvg, Bechtle:2015pma}
which have recently been merged into {\tt HiggsTools}
  \cite{Bahl:2022igd},  both interfaced with $\texttt{ScannerS}$.
$\texttt{HiggsSignals}$ provides a $\chi^2$ value used to
check for the agreement between the predicted observables of the
SM-like Higgs boson in our model and the experimental results.  $\texttt{HiggsBounds}$ checks the parameter points against the exclusion bounds obtained by the experiments in their searches of new scalars. 


As discussed, the N2HDM provides two DM candidates in the FDP, which is the one of interest to us. The DM particle with origin in the singlet has no other restrictions besides the ones from the relic density measurement. The one
with origin in the doublet is compelled to be also in agreement with
direct and indirect detection experiments. For the range of parameters
under study, DD constraints dominate over indirect detection limits and, where applicable, over collider limits. 
The most recent measurements for DD are from the $\texttt{LUX-ZEPLIN}$ (LZ) experiment~\cite{LZ:2022lsv}. 
The results provide a constraint on the spin-independent DM-nucleon direct detection (SI-DD-N) cross section $\hat{\sigma}_\text{SI-DD-N}$. 
This exclusion limit assumes the observed relic density. Since in our scenario the FO relic density is below the measured value, the cross section has to be
multiplied by the corresponding reduced fraction such that the comparison
with the experimental results is performed using
\begin{equation}
    \sigma_\text{SI-DD-N} =
    \hat{\sigma}_\text{SI-DD-N}\frac{\Omega_{\text{FO}}
      h^2}{\Omega_\text{exp}h^2} \; .
    \label{eq:normalizedcxn}
\end{equation}

There is also an indirect constraint on the FO dark sector because of
the existence of a charged scalar $H_D^\pm$.
Since we are dealing with a full dark sector, the couplings of
  $H_\text{SM}$ to the remaining SM particles do not change so that
  its production cross section at the LHC remains unchanged. However,
  the decay width of the SM-like Higgs into photons,
  $H_{\text{SM}}\rightarrow \gamma\gamma$, will change relative to the
  SM due to the additional loop with the dark charged Higgs
  boson. This implies a change of the signal strength
  $\mu_{\gamma\gamma}$ in the photon 
  final state, which is defined as 
  \begin{equation}
    \mu_{\gamma\gamma} = \frac{\sigma(pp\rightarrow H_{\text{SM}}\rightarrow \gamma\gamma)_\text{FDP}}{\sigma(pp\rightarrow H_{\text{SM}}\rightarrow \gamma\gamma)_\text{SM}} = \frac{\sigma_\text{prod,FDP}\mathcal{B}(H_{\text{SM}}\rightarrow \gamma\gamma)_\text{FDP}}{\sigma_\text{prod,SM}\mathcal{B}(H_{\text{SM}}\rightarrow \gamma\gamma)_\text{SM}} =\frac{\mathcal{B}(H_\text{SM}\rightarrow
  \gamma\gamma)_\text{FDP}}{\mathcal{B}(H_\text{SM}\rightarrow
  \gamma\gamma)_\text{SM}} \;,
\end{equation}
Here, $\sigma_\text{prod, i}$ $\ind{\text{SM},\text{FDP}}$ is the
production cross section in either the SM or the new model, i.e.~the FDP of the N2HDM, respectively, and $\mathcal{B}(H_{\text{SM}}\rightarrow \gamma\gamma)_i$ is the corresponding branching ratio of the Higgs decay into two photons. The signal strength is constrained by the latest measurement of the ATLAS experiment \cite{ATLAS:2022tnm}. The measured value at $1\sigma$ is given by
\begin{equation}
    \mu_{\gamma\gamma} = 1.04^{+0.10}_{-0.09}\;.
\end{equation} We demand for the scanned points that the signal
strengths lie in the $2\times 1\sigma$ bound of the measured value,
i.e.~between $0.86$ and $1.24$.

\subsection{Scans}
The parameter scans are performed using $\texttt{ScannerS}$ which checks for all the theoretical and experimental constraints described in the previous section. The relic density and direct detection cross section is calculated with {\tt micrOMEGAs}~\cite{Belanger:2013oya, Belanger:2018mqt}. 
We set the VEV $v$ to the SM value given by
\begin{equation}
    v = \frac{1}{\sqrt{\sqrt{2}G_F}} \simeq 246.22 \, \text{GeV} \;, 
\end{equation}
where $G_F$ denotes the Fermi constant \cite{ParticleDataGroup:2022pth}. One of the
CP-even scalars is the SM-like Higgs boson with the experimentally
measured mass value
\begin{equation}
    m_{H_\text{SM}} = 125.09 \, \text{GeV} \; . 
\end{equation} 
The DM masses from the doublet are varied in the range  
\begin{equation}
    60 \, \text{GeV} \,  \leq \mass{H_{\text{DD}}},\mass{A_{\text{D}}},\mass{H^\pm_{\text{D}}} \leq 1 \, \text{TeV} \; ,
\end{equation}
whereas we vary the mass of $H_\text{DS}$ in the range
\begin{equation}
    1 \, \text{GeV} \,  \leq \mass{H_\text{DS}} \leq  \, 1 \text{TeV}   \; .
\end{equation}
For the FI mechanism to happen the values of $\lambda_7$ and $\lambda_8$ are varied in the range
\begin{equation}\label{couplings}
    10^{-14} \leq \lambda_7, \lambda_8 \leq 10^{-9} \, ,
\end{equation}
while  $\lambda_i$ with $\ind{2,6}$ are varied in 
\begin{equation}
    0 \leq \lambda_2, \lambda_6 \leq 20 \; .
\end{equation}
The scan range for the mass parameter $m_{22}^2$ is given by
\begin{equation}
    0 \leq m_{22}^2 \leq  10^6 \, \text{GeV}^2  \; .
\end{equation}

\subsection{Results}

Once all constraints are imposed we have obtained the sample of points used in the results shown in this section. Let us first state once more
one of the main goals of this work: we are interested in understanding what is the parameter space where FI and FO are complementary and if there is 
something particular to this parameter region from the point of view of phenomenology.
As previously defined, the full relic density $\Omega_{c}h^2$ is the sum of the relic densities $\Omega_\text{FO}$ and $\Omega_\text{FI}$,
\begin{equation}
    \Omega_{c}h^2 = \Omega_\text{FO} h^2 + \Omega_\text{FI}h^2\; .
\end{equation}
\begin{figure}[h!]
    \centering
    \includegraphics[width=7.9cm]{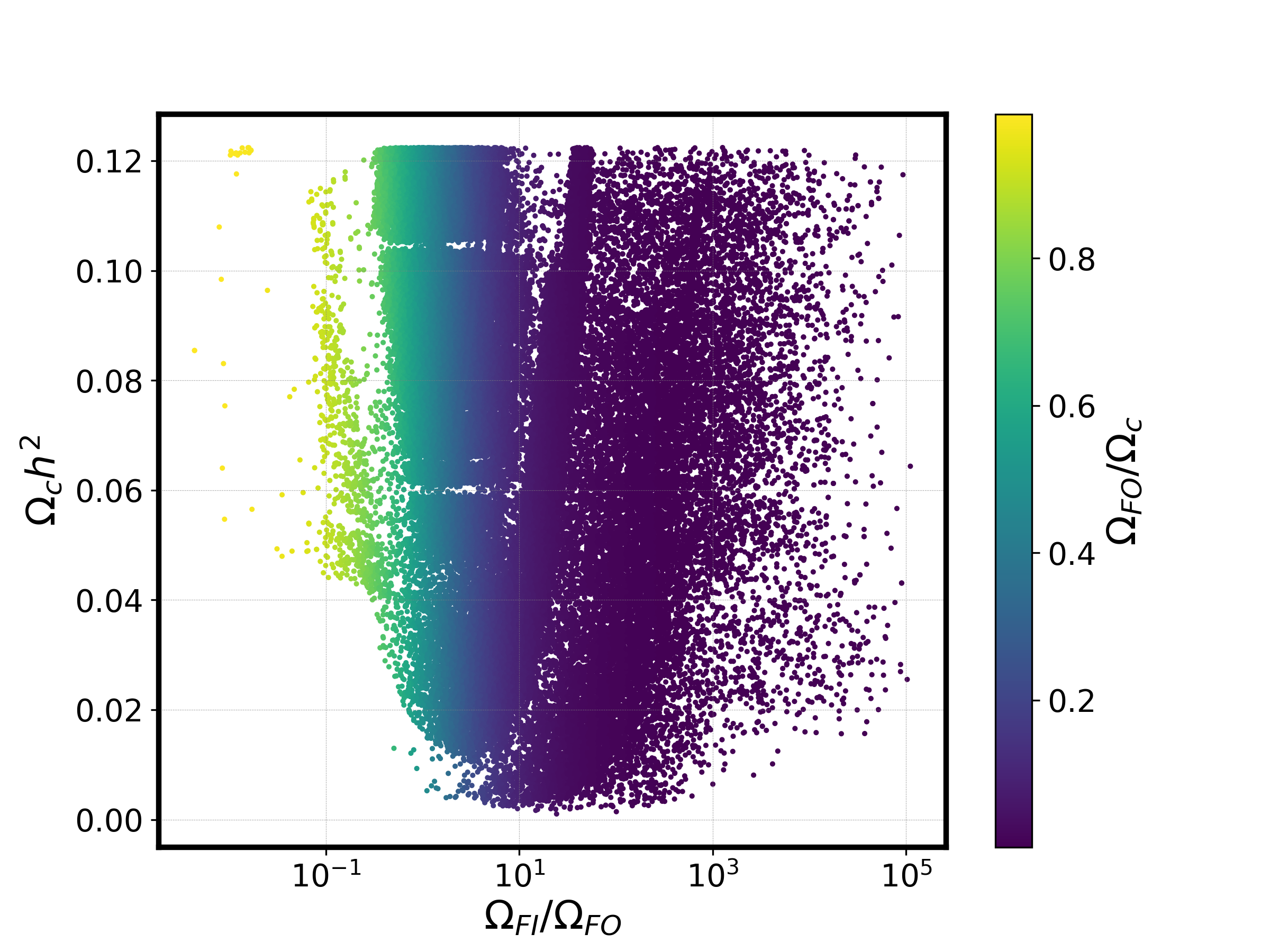}
     \includegraphics[width=7.9cm]{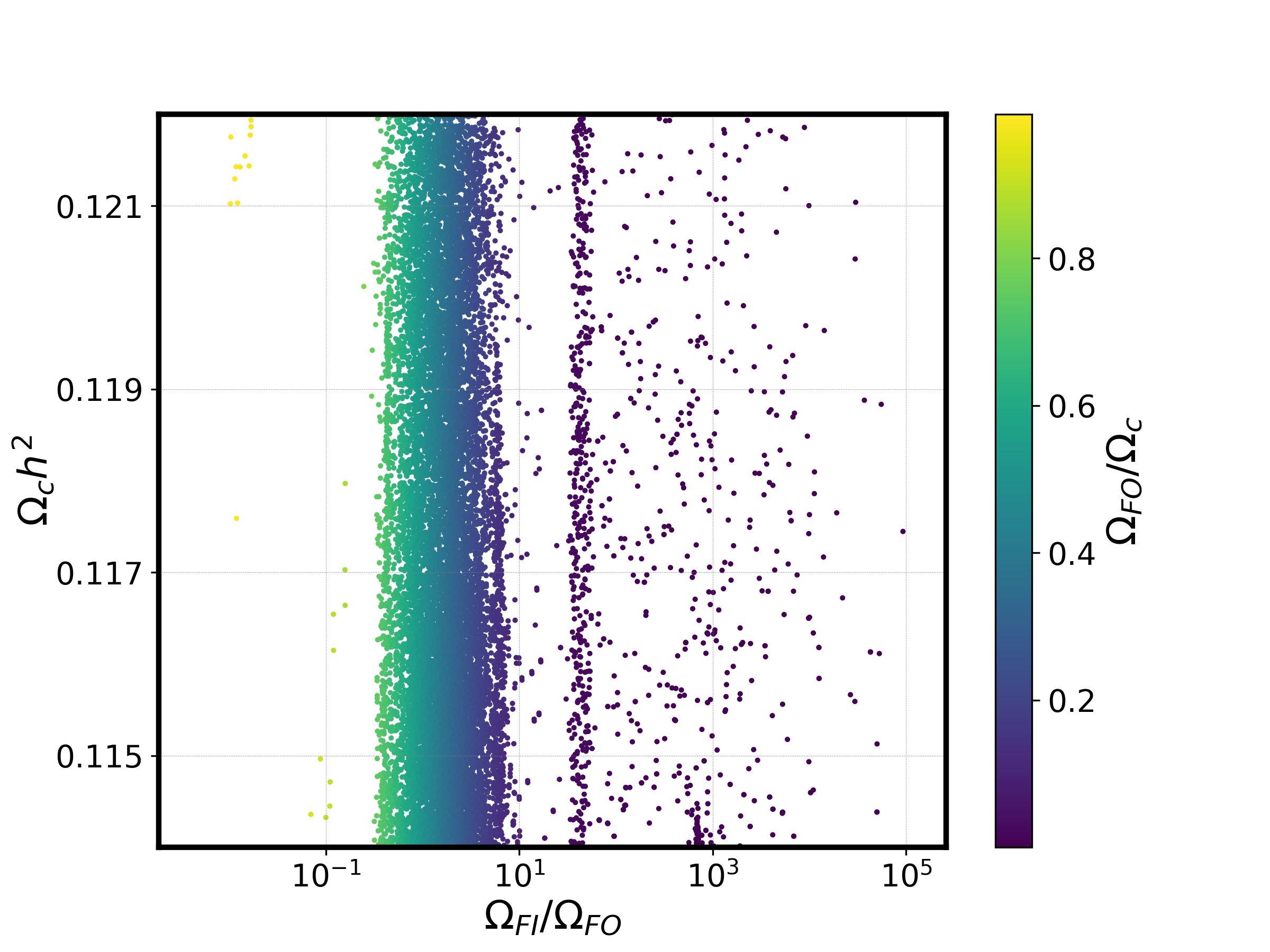}
    \caption{Left: calculated relic density $\Omega_{c}h^2$ as a function of the ratio between the FI relic density $\Omega_{\text{FI}}$ and the FO relic density $\Omega_{\text{FO}}$; 
    the color gradient shows the ratio of the relic density $\Omega_{\text{FO}}$ generated via FO and the full relic density. Right: zoom on the left plot showing the region where the relic density is saturated.}
    \label{fig:relicdiff}
    \end{figure}
In the left plot of Fig.~\ref{fig:relicdiff} we present the full relic density $\Omega_ch^2$ as a function of the ratio between the FI and the FO relic density. The color gradient shows 
the ratio of the FO relic density and the full relic density. 
The right plot is a zoom on the left plot showing in more detail the region where the relic density is saturated.
As expected, most points in the scanned parameter region are dominated by FI. The reason is clear. Once a sector of the model is built such that a DM particle exists with a very small portal coupling, FI is always an option.
If the ratio of $\Omega_{\text{FI}}/\Omega_{\text{FO}}>1$, the contribution of FI to the full relic density is more than $50\%$, whereas if it is less than 1, the contribution of FO is now more than $50\%$. The color gradient provides additional insight by showing the correlation between the relic density $\Omega_{\text{FO}}$ generated through FO and the full relic density. As the ratio $\Omega_{\text{FI}}/\Omega_{\text{FO}}$ decreases, indicating an increase in the importance of FO, FI still plays a crucial role in accounting for the total relic density.  

\begin{figure}[h!]
    \centering
    \includegraphics[width=0.49\linewidth]{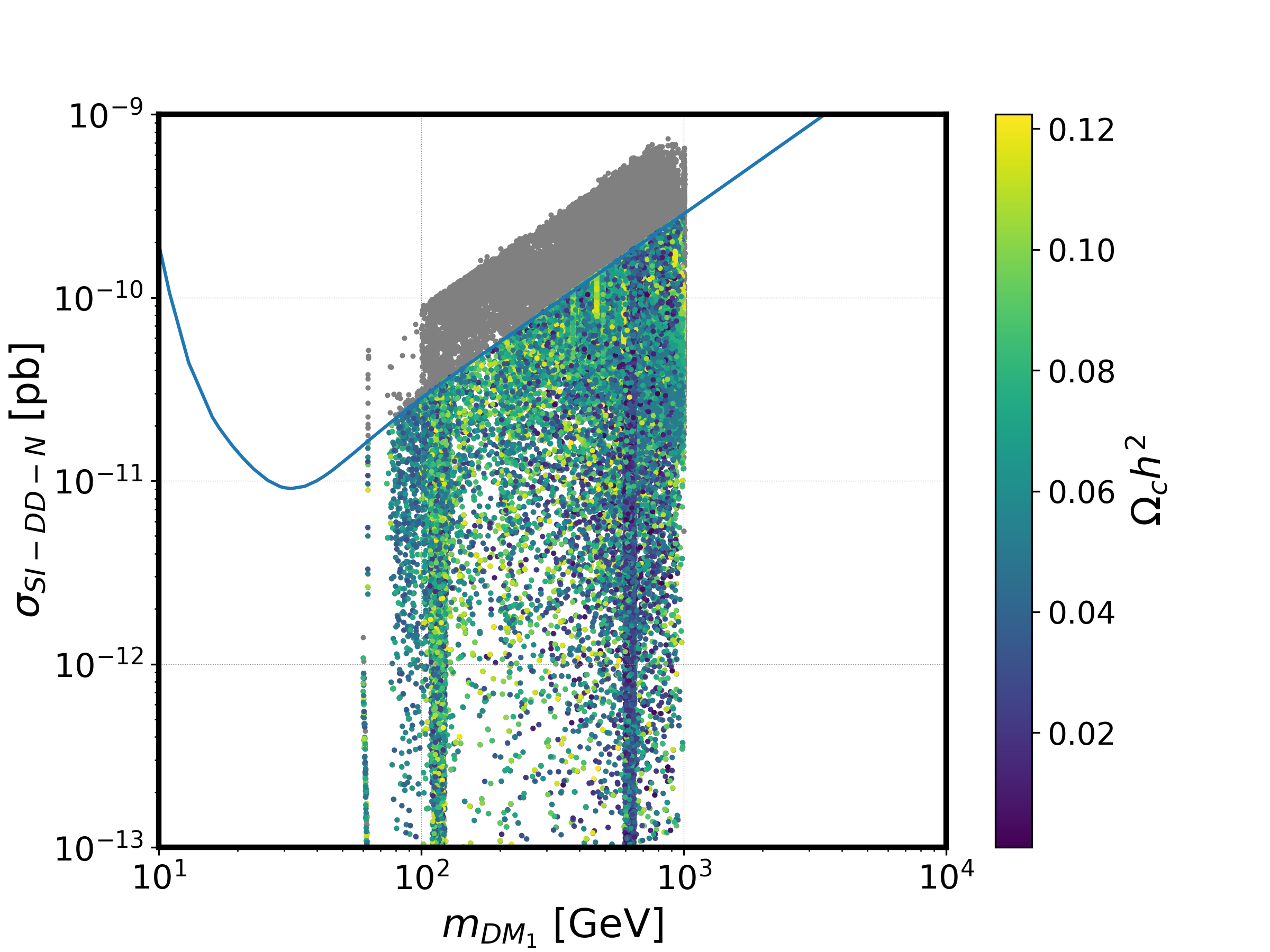}
        \includegraphics[width=0.49\linewidth]{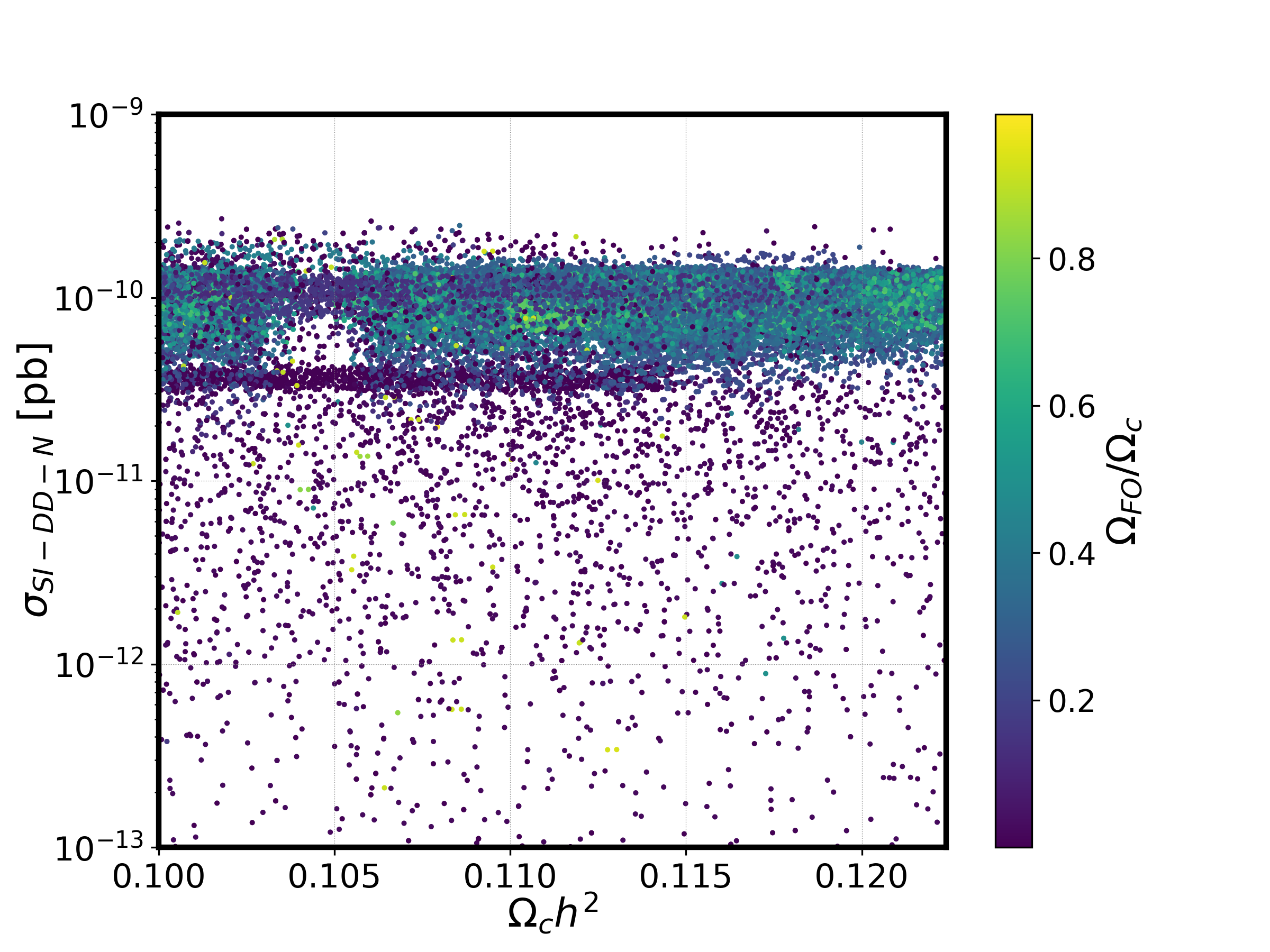}
    \caption{Left: normalized SI direct detection cross section
      $\sigma_\text{SI-DD-N}$ of the FO DM candidate particle as a
      function of its mass $m_{\text{DM}1}$; the blue line depicts the
      fit function of the $\texttt{Lux-Zeplin}$ exclusion bound. The
      gray points are from the full scan while the colored points are
      also allowed by the LZ direct detection experiments and the
      newest $\mu_{\gamma\gamma}$ constraints. The color gradient
        shows the value of the calculated full relic density. Right: correlation between the normalized SI direct detection cross section $\sigma_\text{SI-DD-N}$ of the 
    FO particle and the calculated full relic density with the color gradient indicating the ratio between $\Omega_{\text{FO}}$ and $\Omega_\text{c}$.}
    \label{fig:DD1DM1}
\end{figure}

In Fig.~\ref{fig:DD1DM1} we present in the left plot the normalized
spin-independent direct detection cross section
$\sigma_\text{SI-DD-N}$ (cf.~Eq.~(\ref{eq:normalizedcxn})) as a
function of the mass of the FO DM candidate $m_{\text{DM}1}$. This is the lighter
  of the two neutral dark doublet states, $H_{\text{DD}}$ or
  $A_{\text{D}}$, which we denote DM$_1$, and correspondingly its mass
by $m_{\text{DM}_1}$. The color gradient shows the value of the total
relic density $\Omega_ch^2$. The gray points are all scanned points
surviving all experimental and theoretical constraints while the
colored points, besides the previous constraints, also obey the
LZ constraint and the one
from the signal strength $\mu_{\gamma\gamma}$. In the right plot we
present the correlation between the normalized SI direct detection
cross section of the FO particle and the full relic density, with the
color gradient indicating the ratio between $\Omega_{\text{FO}}$ and
the calculated full relic density $\Omega_c$. The conclusion is that it is possible to find parameter
points that comply with the experimental results and which are above
the neutrino floor~\cite{OHare:2021utq} (which lies below about $10^{-11}-10^{-12}$~pb in the considered mass range). These parameter points can be tested in future DM-nucleon-DD
experiments. 
Also, the importance of FI to saturate the relic density is clear,
since a significant amount of the parameter points corresponding to
the observed relic density have a large contribution from FI. Still,
due to the smaller portal coupling needed for FI, the direct detection
cross section for FI is so small that it is way below the neutrino
floor, which does not allow to probe this type of dark matter particle
in future searches in SI-DD experiments. We should also stress that,
even when FI is the dominant process, implying that
$\Omega_{\text{FO}}/\Omega_c$ has to be small, the SI direct detection
cross section for FO can still be quite large, meaning that it remains
possible to detect the FO DM particle in DD experiments in the
scenario where FI is dominant. This occurs because even though
$\Omega_{\text{FO}}/\Omega_c$ is small, which should decrease the
value of $\sigma_{\text{SI-DD-N}}$, the FO portal coupling must be
larger than what it would normally be without freeze-in so that the
relic density remains saturated, which in turn can increase
$\sigma_{\text{SI-DD-N}}$ overall. This is why on the right plot of
Fig.~\ref{fig:DD1DM1}, several points where FI is the dominant
contribution result in large DD cross sections for the FO
particle. 

As stated above, the loop-induced SM branching ratio into the photons changes
  because of the possibility of the dark charged Higgs running in the loops.
The coupling entering the corresponding loop is $\lambda(H_{\text{SM}}, H_D^\pm, H_D^\pm)$, which is given by
\begin{equation}\label{higgsCont}
    \lambda(H_{\text{SM}}, H_D^\pm, H_D^\pm) = -\lambda_3v\;.
\end{equation}
The signal strength is hence
independent of the couplings $\lambda_7$ and $\lambda_8$, which are
responsible for the FI mechanism. It depends, however, on
  $\lambda_3$ which enters also the couplings of the dark doublet
  states to the SM Higgs boson and hence makes a connection between
  the $\mu_{\gamma\gamma}$ value and the processes contributing to
  direct detection and the relic density of the freeze-out particle. The implications are discussed in the following.

\begin{figure}[h!]
    \centering
        \includegraphics[width=0.49\linewidth]{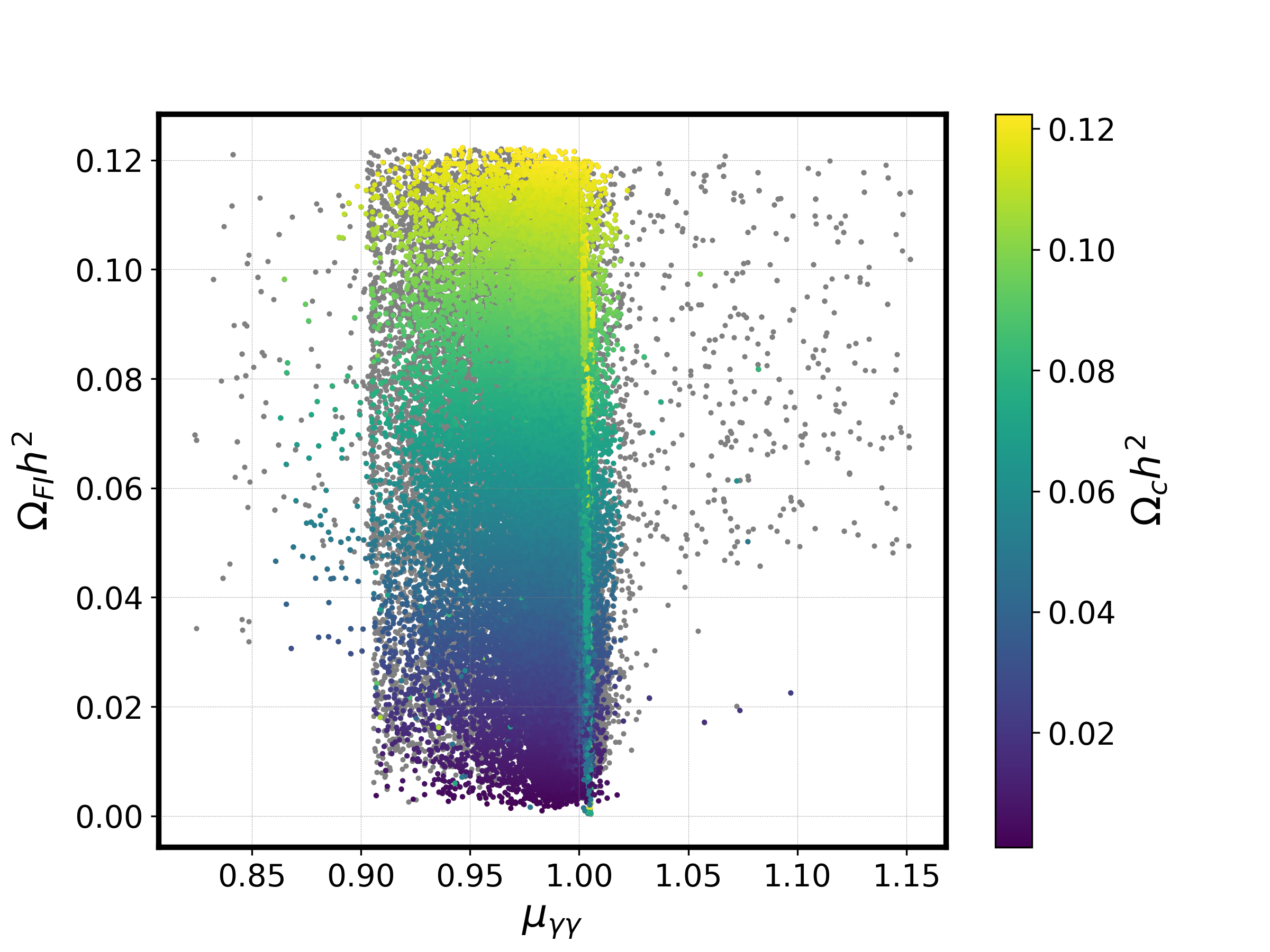}
        \includegraphics[width=0.49\linewidth]{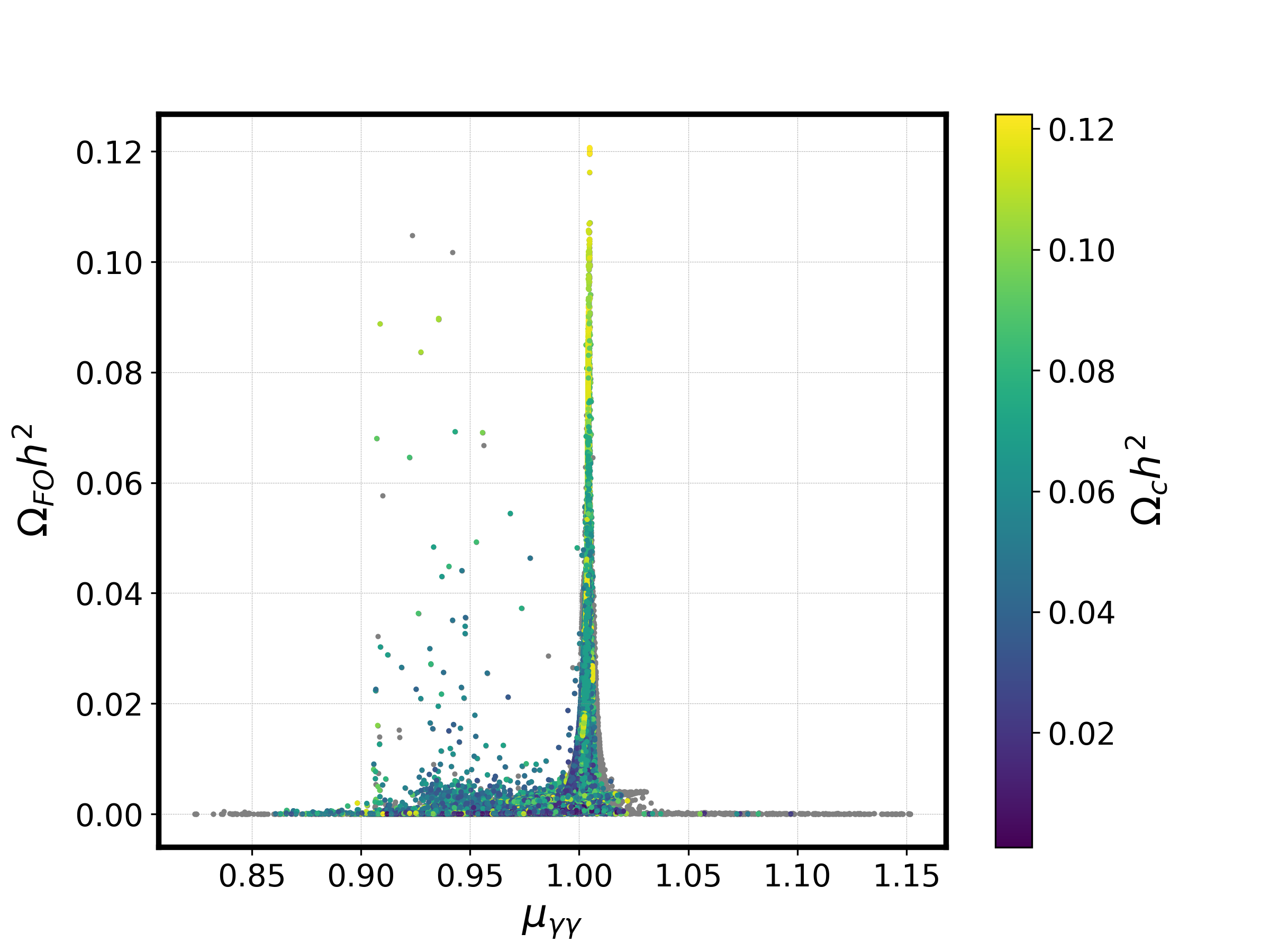}
    \caption{On the left (right), the relic density generated via FI (FO), $\Omega_\text{FI}h^2$ ($\Omega_\text{FO}h^2$), as a function of the signal strength $\mu_{\gamma\gamma}$ is shown. The color gradients show the full relic density $\Omega_ch^2$.}
    \label{DiPhotonFIFO}
\end{figure}

In Fig.~\ref{DiPhotonFIFO} we present the relic densities
$\Omega_\text{FI}h^2$ for FI (left) and $\Omega_\text{FO}h^2$  for FO
(right) as a function of the signal strength $\mu_{\gamma\gamma}$. The
color gradient denotes the full relic density $\Omega_{c}h^2$. The
gray points show the full scan without the updated LZ and
$\mu_{\gamma\gamma}$ constraints. We observe that for all values of
$\mu_{\gamma\gamma}$ below about one, the observed value of the relic
density can be achieved through the FI mechanism. Above one, most
points are excluded by the direct detection constraint of the FO
candidate. As stated above, $\mu_{\gamma\gamma}$
does not depend on the FI 
parameters $\lambda_7$ and $\lambda_8$.  
On the other hand, in the case of FO, the parameter space that
saturates the relic density is close to the SM value since only values
very close to $\mu_{\gamma\gamma} = 1$ are allowed in this
  case. Thus, if we were to measure a $\mu_{\gamma\gamma}$ value
clearly below one, it could hint to the existence of freeze-in in this
model. 

The fact that $\mu_{\gamma\gamma}\approx 1$ for FO provides insight
into the parameter regions where the FO mechanism becomes the dominant
process. In fact, only if the additional contribution of the charged
Higgs $H^\pm_D$ loop is small, we can have a SM-like value of
$\mu_{\gamma\gamma} \approx 1$. This can be achieved if $\lambda_3$ is small due to the dependence of $H_\text{SM} H^\pm_\text{D} H^\pm_\text{D}$ on that parameter only. The parameter  $\lambda_3$ can be written as a function of the input parameters $m^2_{H^\pm_D}$ and $m^2_{22}$ as
\begin{equation}
    \lambda_3 = \frac{2(m_{H^\pm_D}^2-m^2_{22})}{v^2} \; .
\end{equation}
\begin{figure}[h!]
    \centering
        \includegraphics[width=0.49\linewidth]{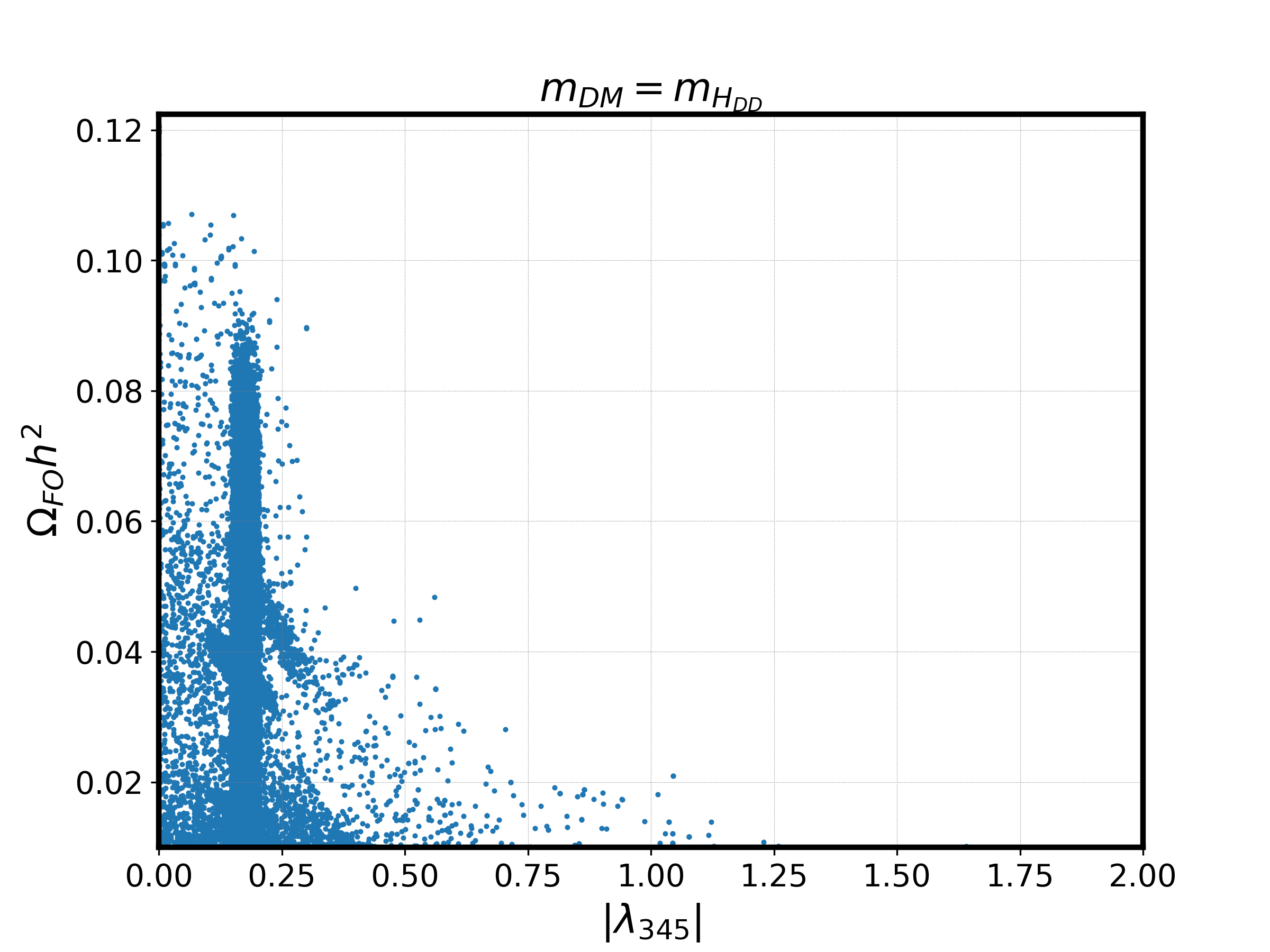}
        \includegraphics[width=0.49\linewidth]{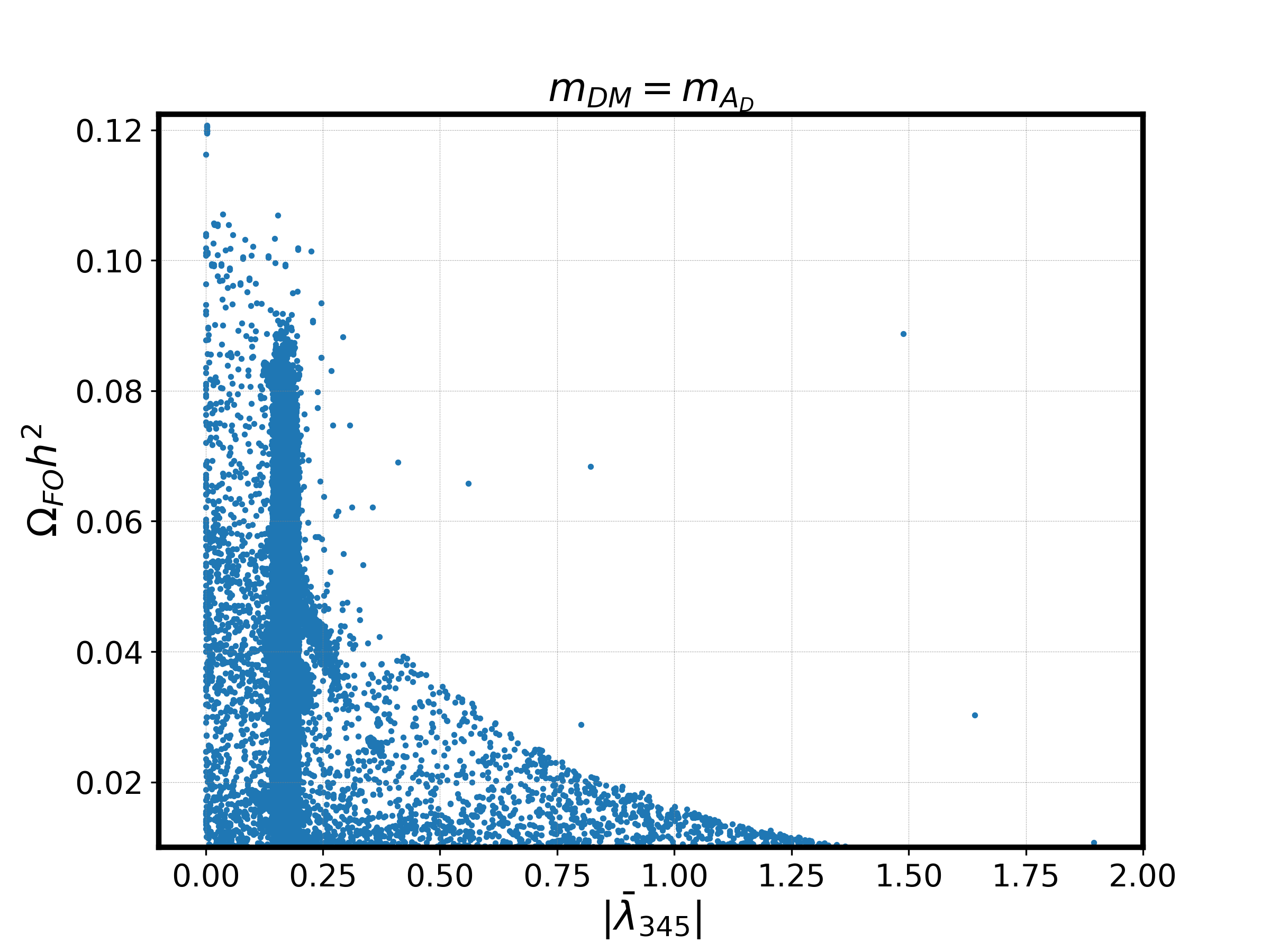}
    \caption{On the left we present the relic density generated via FO as a function of the absolute value of $\lambda_{345}$. On the right we show the FO relic density as a function of the absolute value of $\bar{\lambda}_{345}$.}
    \label{FreezeOutDOM}
\end{figure} 
Consequently, to make $\lambda_3$ small, one needs to choose
$m^2_{H^\pm_D} \approx m^2_{22}$.
On the other hand, a small value of $\lambda_3$ is
  directly related to a significant value of the FO relic density, which
  can be explained as follows. The TAC for freeze-out is forced to be
small in order to obtain a large FO relic density. The TAC depends on
the Higgs portal couplings and the quartic couplings.\footnote{The
  vertices, where these couplings appear, are given in Appendix \ref{FDPApp}.} The relevant couplings involving $H^\pm_\text{D}$ are proportional to $\lambda_3$. Therefore, choosing a small $\lambda_3$ leads to a lower TAC and hence a higher final relic density. However, choosing $\lambda_3$ to be small is not the only condition for a large FO contribution to the final relic density.
In Fig.~\ref{FreezeOutDOM} we show the FO relic density as a function
of the absolute value of the coupling combinations $|\lambda_{345}|=
|\lambda_3 + \lambda_4 +\lambda_5|$ (left plot) and
$|\Bar{\lambda}_{345}|=|\lambda_3 +\lambda_4 - \lambda_5|$  (right
plot). One can see that small values of these coupling combinations
lead to a large FO relic density. The coupling combination
$\lambda_{345}$ appears in the Higgs portal coupling
$\lambda(\HD,\HD,\SM)$ and the quartic coupling
$\lambda(\HD,\HD,\SM,\SM)$, and the combination $\bar{\lambda}_{345}$
appears in the couplings $\lambda(\AD,\AD,\SM)$ and
$\lambda(\AD,\AD,\SM,\SM)$.  The conditions for a large FO relic density are
therefore that $\lambda_3$ and the coupling combinations
$\lambda_{345}$ or $\Bar{\lambda}_{345}$ (depending on whether
$H_{\text{DD}}$ or $A_{\text{D}}$ is the DM FO particle) must be
small. This in turn explains why a significant FO relic density is
directly related to a photonic signal strength close to 1. The FI mechanism is needed to obtain the observed relic density if neither of these conditions are fulfilled.

\subsection{A Note on CP-violation in the FI+FO Framework}

Starting with the scalar potential of the FDP of the N2HDM, if instead of two symmetries $\mathbb{Z}^{(1)}_{2}$ and $\mathbb{Z}^{(2)}_{2}$ we impose just one but different symmetry
\begin{equation}
\mathbb{Z}^{(3)}_{2}:\quad \Phi_{1}\rightarrow \Phi_{1},\quad \Phi_{2}\rightarrow -\Phi_{2},\quad \Phi_{S}\rightarrow - \Phi_{S}\,,\label{eq:Z2_3}
\end{equation}
the most general renormalisable scalar potential invariant under this symmetry can be written as
\begin{equation}
V_{\text{CPD}} = V_{\text{N2HDM}}^{\text{FDP}}  + (A
\Phi_{1}^{\dagger} \Phi_{2}  \Phi_{S} + \text{h.c.}) \;,
\end{equation}
which was dubbed in  \cite{Azevedo:2018fmj} as CP in the dark
(CPD).\footnote{The model was investigated w.r.t.~its possibility of a strong first-order phase transition in \cite{Biermann:2022meg}.} The extra term allows for CP-violation in the dark sector. As
shown in~\cite{Azevedo:2018fmj}, FO is possible in a substantial region of the parameter space. The question is now if FI is also possible in the model, and if the two processes can occur simultaneously. We note that in this model 
all dark neutral states couple to gauge bosons. 

In order to understand the problem let us start by defining the eigenstates of the neutral dark sector as $h_i$, with $i=1,2,3$. The mass eigenstates can be obtained from the gauge eigenstates defined in Eq.~(\ref{eq:phis}) 
via the orthogonal rotation matrix  parameterised by the angles $\alpha_1$, $\alpha_2$ and $\alpha_3$ with $\alpha_i\in \left[-\frac{\pi}{2},\frac{\pi}{2}\right]$,
\begin{equation}
    R=\begin{pmatrix}
    c_{\alpha_1}c_{\alpha_2}&s_{\alpha_1}c_{\alpha_2}&s_{\alpha_2}\\
        -(c_{\alpha_1}s_{\alpha_2}s_{\alpha_3}+s_{\alpha_1}c_{\alpha_3})&c_{\alpha_1}c_{\alpha_3}-s_{\alpha_1}s_{\alpha_2}s_{\alpha_3}&c_{\alpha_2}s_{\alpha_3}\\
        -c_{\alpha_1}s_{\alpha_2}c_{\alpha_3}+s_{\alpha_1}s_{\alpha_3}&-(c_{\alpha_1}s_{\alpha_3}+s_{\alpha_1}s_{\alpha_2}c_{\alpha_3})&c_{\alpha_2}c_{\alpha_3}
    \end{pmatrix}\;,\label{rotmat}
\end{equation}
where the notation $\sin(\alpha_i)\equiv s_{\alpha_i}$ and $\cos(\alpha_i)\equiv c_{\alpha_i}$ was used. 

The first step is to look at the coupling strengths between the dark sector particles and the SM. For the particles that freeze-out, 
the coupling strength has to be at least of the order $10^{-3}$, while for the particles that FI it has to be below about $10^{-8}$. The only way to achieve such a low value is to force one of the $h_i$ to decouple from the $Z$ and $W$ bosons 
since the $SU(2)_L$ gauge coupling is of order $10^{-1}$.
Let us take $h_1$ to be the state to decouple from the gauge bosons which in turn means that the absolute value of the couplings of the vertices $Zh_1h_j$, which can be derived from the kinetic terms for $\Phi_2$,  
\begin{align}
    \left| D_\mu \Phi_2 \right|^2\ni& \frac{-g}{cos\theta_W}c_{\alpha_2}c_{\alpha_3} Z_\mu\left( h_1\partial^\mu h_2-h_2\partial^\mu h_1 \right)\hspace{3mm}\mathrm{for}\hspace{1mm}j=2\hspace{2mm}\mathrm{and}\\
    \left| D_\mu \Phi_2 \right|^2\ni& \frac{g}{cos\theta_W}c_{\alpha_2}s_{\alpha_3} Z_\mu\left( h_1\partial^\mu h_3-h_3\partial^\mu h_1 \right)\hspace{3mm}\mathrm{for}\hspace{1mm}j=3\;,
\end{align}
have to be very small. If we choose $\alpha_2\rightarrow \frac{\pi}{2}$ the rotation matrix becomes
\begin{equation}
    R=\begin{pmatrix}
    0&0&1\\
    -s_{\alpha_1+\alpha_3}&c_{\alpha_1+\alpha_3}&0\\
    -c_{\alpha_1+\alpha_3}&-s_{\alpha_1+\alpha_3}&0
    \end{pmatrix}\hspace{3mm}\label{h1R}
\end{equation}
and the mass eigenstates
\begin{align}
    h_1=&\rho_s\nonumber \, , \\
    h_2=&-s_{\alpha_1+\alpha_3}\rho_2+c_{\alpha_1+\alpha_3}\eta_2 \, , \\
    h_3=&-c_{\alpha_1+\alpha_3}\rho_2-s_{\alpha_1+\alpha_3}\eta_2\nonumber \, ,
\end{align} 
and $h_1$ becomes a singlet like field and can therefore be the FI DM candidate. However, when we decouple one of the scalars we lose CP-violation in the scalar sector because we need to set $A=0$ and $\lambda_5 = 0$. This can also be seen by their definitions via the remaining parameters
\begin{align}
m_{h_3}^2=&-\frac{m_{h_2}^2R_{21}R_{22}+m_{h_1}^2R_{11}R_{12}}{R_{31}R_{32}}=m_{h_2}^2\label{mh3par}\,
            ,\\\nonumber\\
\lambda_5=&\frac{R_{13} (m_{h_3}^2 - m_{h_1}^2 R_{23}^2 + m_{h_2}^2
            (R_{23}^2-1) + (m_{h_1}^2 - m_{h_3}^2) R_{33}^2)}{v^2
            (R_{13} - 2 R_{21} R_{32})}=0\label{l5par}\, ,\\\nonumber\\
\mathrm{Re}(A)=&\frac{R_{11} ((m_{h_2}^2 - m_{h_1}^2) R_{21}^2 +
                 (m_{h_1}^2 - m_{h_3}^2)R_{31}^2)}{v^2 (R_{13}-2
                 R_{21} R_{32})}=0\label{realA}\, ,\\\nonumber\\
\mathrm{Im}(A)=&\frac{R_{12} ((m_{h_2}^2 - m_{h_1}^2) R_{21}^2 +
                 (m_{h_1}^2 - m_{h_3}^2)R_{31}^2)}{v^2 (R_{13}-2
                 R_{21} R_{32})}=0\label{imagA} \, ,
\end{align}
with  the components $R_{ij}$ of the rotation matrix given in Eq.~(\ref{h1R}).

 These couplings can be made zero by evoking two symmetries: the one
 already imposed, $\mathbb{Z}^{(3)}_{2}$, and a $U(1)$ symmetry under
 which $\Phi_1 \to \Phi_1, \, \, \Phi_2 \to e^{i \theta} \Phi_2, \,
\, \Phi_S \to  \Phi_S$. In this restricted version of the model we can
now have FI for $h_1$ and FO for $h_{2,3}$ (with the same mass). The
model is now very similar to the one previously discussed
and we will just show that the results are similar.

\begin{figure}[h!]
  \centering
\includegraphics[width=0.49\linewidth]{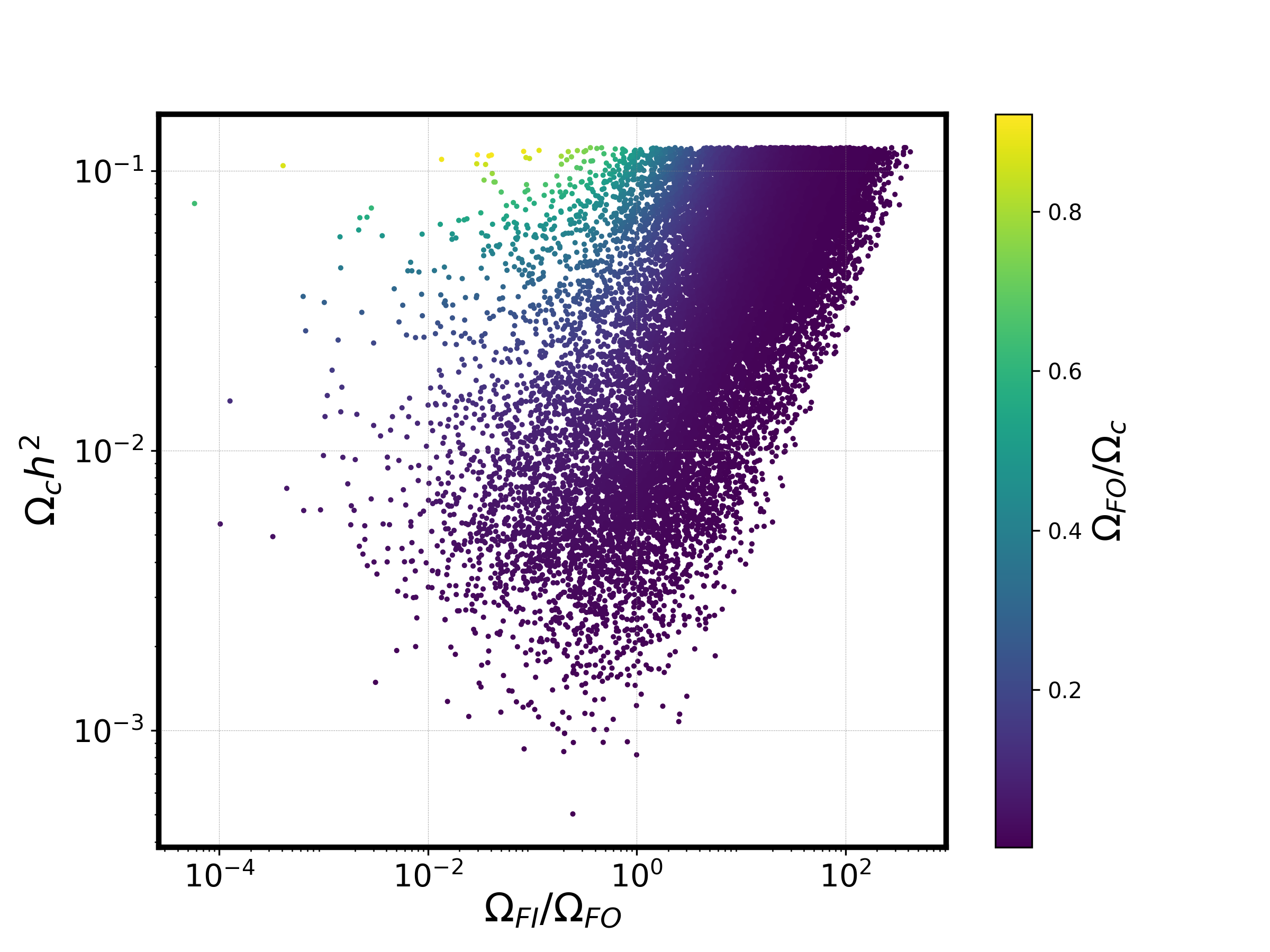}
\includegraphics[width=0.49\linewidth]{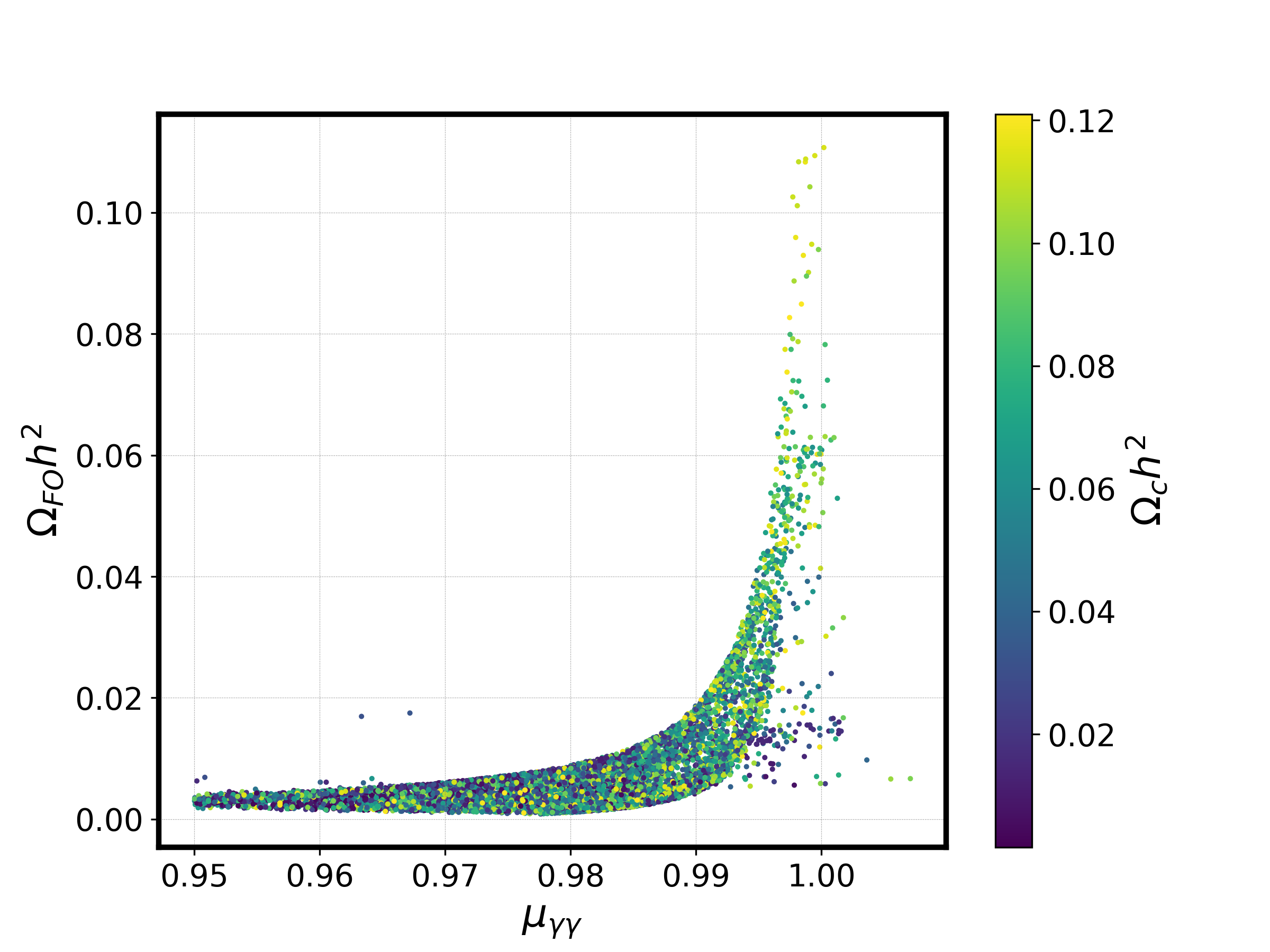}
  \caption{Left: calculated relic density  as a function of the ratio between the FI relic density and the FO relic density; 
    the color gradient shows the ratio of the relic density generated
    via FO and the total relic density. Right: Freeze-out relic
density as a function of $\mu_{\gamma\gamma}$; the color gradient
shows the calculated relic density.
    }\label{fig:aa}
\end{figure}
In Fig.~\ref{fig:aa} we show in the left plot the calculated relic
density  as a function of the ratio between the FI relic density and
the FO relic density. The color gradient shows the ratio of the relic
density generated via FO and the total relic density. As for the
N2HDM, the full relic density has most of the times a much larger
contribution from FI. In the right plot we
  present the freeze-out relic
density as a function of $\mu_{\gamma\gamma}$. The color gradient
shows the calculated relic density. Here again, and for similar reasons, the region where FO has a large contribution coincides with a  $\mu_{\gamma \gamma}$ close to one.

\begin{figure}[h!]
  \centering
 \includegraphics[width=0.49\linewidth]{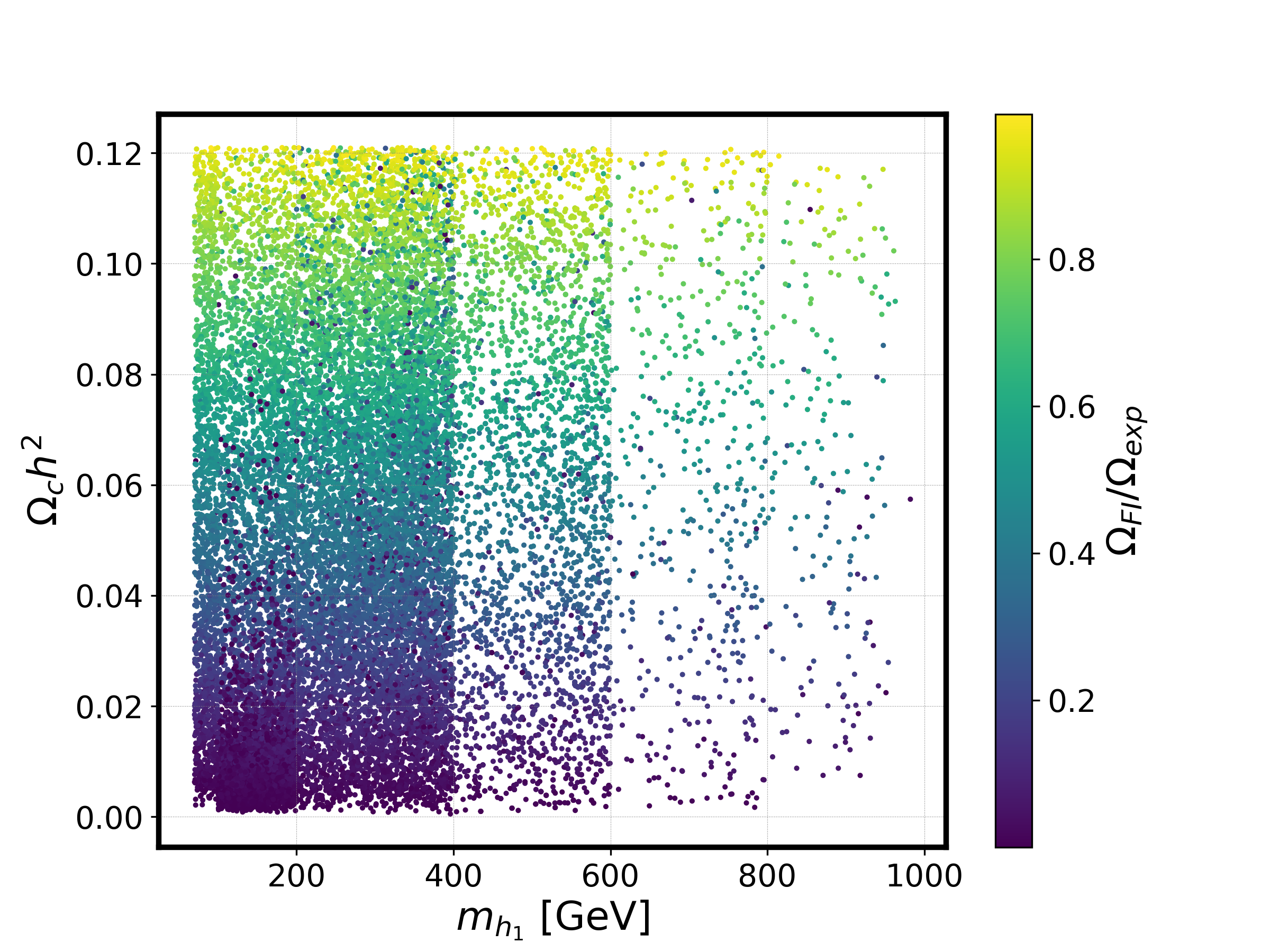}
\includegraphics[width=0.49\linewidth]{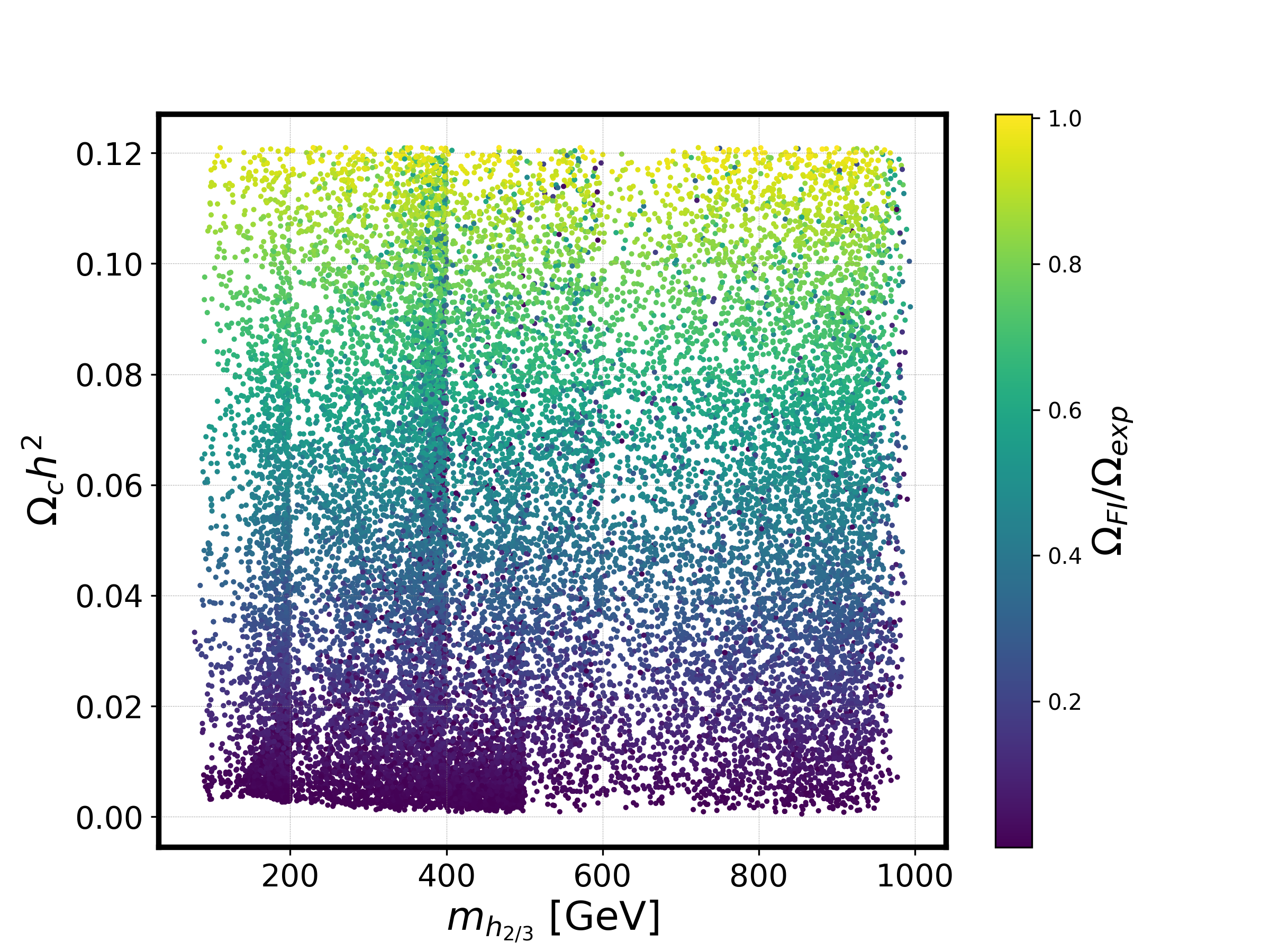}
  \caption{Calculated relic density  as a function of the FI mass (left) and FO mass (right);
    the color gradient shows the ratio of the relic density generated via FI and the observed relic density. 
    }\label{fig:bb}
\end{figure}
In Fig.~\ref{fig:bb} we present the calculated relic density as a function of the FI mass (left) and the FO mass (right). The color gradient shows the ratio of the relic density generated via FI and the observed relic density. The important point to note is that the entire region for the DM FO masses in the scan is allowed. Also, we see again that in order to obtain the observed relic density, the contribution from the FI candidate is larger than the one from the FO candidates for the majority of points, and without FI, most of the points would be excluded. Although for most of the parameter space the FO candidates contribute to only a fraction of the total relic density, they could still be detected both in direct detection and collider experiments.

In summary, CP in the Dark is able to generate the observed relic density via the two complementary mechanisms of FO and FI. However, this is achieved at the cost of losing the CP-violation feature in the dark scalar sector. If we leave the CP in the dark model as it is (i.e. not choosing the $\alpha_2=\pi/2$) we keep the CP violating properties. From here we add another singlet to the model which is then able to fill the relic density via FI. This would allow for a scenario with a strong first order electroweak phase transition and could possibly explain the baryon asymmetry of the universe.

\section{Conclusions \label{sec:conclusions}}

In this work we have discussed the possibility of having two DM candidates, one produced via freeze-in and the other via freeze-out. We have shown that even a simple extension
with only two extra singlets is     able to implement the idea. We have then shown that also other extensions such as the N2HDM in the Full Dark Phase and the CP in the Dark model, 
have this complementarity between freeze-in and freeze-out if two independent $Z_2$ symmetries are imposed. Hence, any model that does not fulfil the relic density can be easily extended with a new field and a new symmetry that stabilises it such that it works via freeze-in.

It is clear that freeze-in is always possible but has the problem of being hard to probe although in some scenarios it could be feasible (see for instance~\cite{Belanger:2020npe, Agrawal:2021dbo}). Now, when we see the processes as complementary, the possible values of the freeze-out portal coupling vary in a wider range, allowing for a larger portal coupling. This in turn will make searches at the LHC much more interesting while the relic density is still saturated by the freeze-in contribution. At the same time, the dark matter fraction can be small enough that the FO DM candidate cannot be detected via direct and indirect detection but could be probed at colliders. Nevertheless, we should emphasise that even for a small FO density fraction, the SI-DD cross section can be quite large, as we have shown for the FDP of the N2HDM and for CPD. This in turn means that it can still happen that both collider and direct detection and indirect detection experiments are sensitive to the FO particle even when freeze-in is the dominant process.

If a DM particle is found in a direct detection experiment this complementarity should always be considered, especially if a proposed model does not match the measured relic density and/or an unexpected result at a collider is found. This also signifies that direct detection cannot be used as a guide to exclude searches in regions of the parameter space of a given model and that searches at the LHC have to disregard the bounds from DD in particular scenarios. 

As more data is gathered in direct and 
indirect detection and collider experiments, the constraints on the DM mass and portal couplings will become more stringent, which may provide clues as to which models and mechanisms can explain DM.

\vspace*{1cm}
\appendix
\section{Full Dark Phase of the N2HDM Couplings}\label{FDPApp}
\input{Appendix2.tex}

\subsubsection*{Acknowledgments}

RC and RS are partially supported by the Portuguese Foundation for Science and Technology (FCT) under Contracts no. UIDB/00618/2020, UIDP/00618/2020, CERN/FIS-PAR/0025/2021 and CERN/FIS-PAR/0021/2021. RC is additionally supported by FCT with a PhD Grant No. 2020.08221.BD. KE is grateful to Avicenna Studienstiftung for financial support. JP acknowledges financial support from the Studienstiftung des Deutschen Volkes. MM is grateful for support from the Deutsche Forschungsgemeinschaft (DFG, German Research Foundation) under grant 396021762 - TRR 257.
\\

\vspace*{1cm}
\bibliographystyle{h-physrev}
\bibliography{finfout.bib}

\end{document}

%% file: Appendix2.tex
The coupling strengths $\lambda$ between the dark sector particles of the Full Dark Phase of the N2HDM and the SM particles, and between the dark sector particles themselves\footnote{Note that we do not give the Lorentz part for the gauge couplings, as we are only interested in the prefactors.} are given here. We start with the cubic and quartic interactions between the SM-like Higgs and the dark scalars and the interactions just between dark scalars: 
\begin{align}
\lambda(\SM, \DS, \DS) &= - \lambda_7 v  \\ 
\lambda(\SM, \SM, \DS, \DS) &= -\lambda_7 \\ 
\lambda(\HD, \HD, \DS, \DS) &=\lambda(\AD, \AD, \DS, \DS)=\lambda(H_\text{D}^{+}, H_\text{D}^{-}, \DS, \DS) = -\lambda_8 \\
\lambda(\DS, \DS, \DS, \DS)&= -3\lambda_6\\
\lambda(\SM, \HD, \HD) &= -\lambda_{345}v \label{HD3} \\ 
\lambda(\SM,\SM , \HD, \HD) &= -\lambda_{345}\label{HD4}\\
\lambda(\AD, \AD, \HD, \HD) &=  \lambda(H_\text{D}^{+},H_\text{D}^{-},\HD,\HD)=-\lambda_2 \\ 
\lambda(\HD, \HD, \HD, \HD) &=-3\lambda_2 \\
\lambda(\SM , \AD, \AD) &= -\bar{\lambda}_{345}v\label{AD3}\\ 
\lambda(\SM,\SM , \AD, \AD) &= -\bar{\lambda}_{345}\label{AD4}\\  
\lambda(H_\text{D}^{+},H_\text{D}^{-},\AD,\AD)&=-\lambda_2 \\
\lambda(\AD, \AD, \AD, \AD)&=-3\lambda_2 \\
\lambda(\SM,H_\text{D}^{+},H_\text{D}^{-}) &=-\lambda_3v \label{HP3}\\
\lambda(\SM,\SM,H_\text{D}^{+},H_\text{D}^{-}) &=-\lambda_3 \label{HP4} \\
\lambda(H_\text{D}^{+},H_\text{D}^{+},H_\text{D}^{-},H_\text{D}^{-}) &= -2\lambda_2 \; .
\end{align}

Here, we give the cubic and quartic interactions between the gauge bosons and the dark scalars:
\begin{align} 
\lambda(\HD, H^{+}_\text{D}, W^-) &= - \lambda(\HD, H^{-}_\text{D}, W^+) = \frac{g}{2}\\
\lambda(\HD, \AD, Z) &=-\frac{i}{2}(g\cos{\theta_W}+g'\sin{\theta_W})  \\
\lambda(\HD, \HD, W^+, W^-) &= \frac{g^2}{2}\\ 
\lambda(\HD, H_\text{D}^+, W^-, \gamma) &= \lambda(\HD, H_\text{D}^-, W^+, \gamma) = \frac{1}{2}gg'\cos{\theta_W} \\
\lambda(\HD, H_\text{D}^+, W^-, Z) &= \lambda(\HD, H_\text{D}^-, W^+, Z) = -\frac{1}{2}gg'\sin{\theta_W}\\
\lambda(\HD, \HD, Z, Z) &= \frac{(g\cos{\theta_W}+g'\sin{\theta_W)^2}}{2} \\   
\lambda(\AD, H^+_\text{D}, W^-)&= \lambda(\AD, H^-_\text{D}, W^+) = \frac{i}{2} g\\
\lambda(\AD, \AD, W^+, W^-) &= \frac{g^2}{2} \\
\lambda(\AD, H_\text{D}^{+}, W^{-}, \gamma) &= -\lambda(\AD, H_\text{D}^{-}, W^{+}, \gamma) = \frac{i}{2}gg'\cos{\theta_W} \\
\lambda(\AD, H_\text{D}^+, W^-, Z) &= -\lambda(\AD, H_\text{D}^-, W^+, Z) =  -\frac{i}{2}gg'\sin{\theta_W} \\
\lambda(\AD, \AD, Z, Z) &= \frac{(g\cos{\theta_W}+g'\sin{\theta_W)^2}}{2} \\ 
\lambda(H_\text{D}^+, H_\text{D}^-, \gamma)&=-\frac{1}{2}(g'\cos{\theta_W}+g\sin{\theta_W}) \\
\lambda(H_\text{D}^+, H_\text{D}^-, Z)&=-\frac{1}{2}(-g'\sin{\theta_W} + g\cos{\theta_W}) \\
\lambda(H_\text{D}^{+},H_\text{D}^{-},W^+,W^-) &= \frac{g^2}{2} \\
\lambda(H_\text{D}^{+},H_\text{D}^{-},\gamma,\gamma) &= \frac{1}{2}(g'\cos{\theta_W}+g\sin{\theta_W})^2 \\
\lambda(H_\text{D}^{+},H_\text{D}^{-},\gamma,Z) &= -\frac{1}{4} (-2g'g\cos{2\theta_W} + (g'^2 - g^2)\sin{2\theta_W}) \\
\lambda(H_\text{D}^{+},H_\text{D}^{-}, Z, Z) &= \frac{1}{2}(g\cos{\theta_W} - g'\sin{\theta_W})^2 \; .
\end{align}

%% file: finfout.bbl
\begin{thebibliography}{10}

\bibitem{Patt:2006fw}
B.~Patt and F.~Wilczek,
\newblock (2006), hep-ph/0605188.

\bibitem{Planck:2018vyg}
Planck, N.~Aghanim {\em et~al.},
\newblock Astron. Astrophys. {\bf 641}, A6 (2020), 1807.06209,
\newblock [Erratum: Astron.Astrophys. 652, C4 (2021)].

\bibitem{Zeldovich:1965gev}
Y.~b. Zeldovich,
\newblock Adv. Astron. Astrophys. {\bf 3}, 241 (1965).

\bibitem{Bertone:2004pz}
G.~Bertone, D.~Hooper, and J.~Silk,
\newblock Phys. Rept. {\bf 405}, 279 (2005), hep-ph/0404175.

\bibitem{Hall:2009bx}
L.~J. Hall, K.~Jedamzik, J.~March-Russell, and S.~M. West,
\newblock JHEP {\bf 03}, 080 (2010), 0911.1120.

\bibitem{DEramo:2010keq}
F.~D'Eramo and J.~Thaler,
\newblock JHEP {\bf 06}, 109 (2010), 1003.5912.

\bibitem{Hochberg:2014dra}
Y.~Hochberg, E.~Kuflik, T.~Volansky, and J.~G. Wacker,
\newblock Phys. Rev. Lett. {\bf 113}, 171301 (2014), 1402.5143.

\bibitem{Kuflik:2015isi}
E.~Kuflik, M.~Perelstein, N.~R.-L. Lorier, and Y.-D. Tsai,
\newblock Phys. Rev. Lett. {\bf 116}, 221302 (2016), 1512.04545.

\bibitem{DAgnolo:2015ujb}
R.~T. D'Agnolo and J.~T. Ruderman,
\newblock Phys. Rev. Lett. {\bf 115}, 061301 (2015), 1505.07107.

\bibitem{Pappadopulo:2016pkp}
D.~Pappadopulo, J.~T. Ruderman, and G.~Trevisan,
\newblock Phys. Rev. D {\bf 94}, 035005 (2016), 1602.04219.

\bibitem{Garny:2017rxs}
M.~Garny, J.~Heisig, B.~L\"ulf, and S.~Vogl,
\newblock Phys. Rev. D {\bf 96}, 103521 (2017), 1705.09292.

\bibitem{DAgnolo:2020mpt}
R.~T. D'Agnolo, D.~Liu, J.~T. Ruderman, and P.-J. Wang,
\newblock JHEP {\bf 06}, 103 (2021), 2012.11766.

\bibitem{Smirnov:2020zwf}
J.~Smirnov and J.~F. Beacom,
\newblock Phys. Rev. Lett. {\bf 125}, 131301 (2020), 2002.04038.

\bibitem{Fitzpatrick:2020vba}
P.~J. Fitzpatrick, H.~Liu, T.~R. Slatyer, and Y.-D. Tsai,
\newblock Phys. Rev. D {\bf 106}, 083517 (2022), 2011.01240.

\bibitem{Kramer:2020sbb}
E.~D. Kramer, E.~Kuflik, N.~Levi, N.~J. Outmezguine, and J.~T. Ruderman,
\newblock Phys. Rev. Lett. {\bf 126}, 081802 (2021), 2003.04900.

\bibitem{Hryczuk:2021qtz}
A.~Hryczuk and M.~Laletin,
\newblock JHEP {\bf 06}, 026 (2021), 2104.05684.

\bibitem{Bringmann:2021tjr}
T.~Bringmann, P.~F. Depta, M.~Hufnagel, J.~T. Ruderman, and K.~Schmidt-Hoberg,
\newblock Phys. Rev. Lett. {\bf 127}, 191802 (2021), 2103.16572.

\bibitem{LZ:2022lsv}
LZ, J.~Aalbers {\em et~al.},
\newblock Phys. Rev. Lett. {\bf 131}, 041002 (2023), 2207.03764.

\bibitem{XENON:2023cxc}
XENON, E.~Aprile {\em et~al.},
\newblock Phys. Rev. Lett. {\bf 131}, 041003 (2023), 2303.14729.

\bibitem{McDonald:1993ex}
J.~McDonald,
\newblock Phys. Rev. D {\bf 50}, 3637 (1994), hep-ph/0702143.

\bibitem{Deshpande:1977rw}
N.~G. Deshpande and E.~Ma,
\newblock Phys. Rev. D {\bf 18}, 2574 (1978).

\bibitem{Arhrib:2013ela}
A.~Arhrib, Y.-L.~S. Tsai, Q.~Yuan, and T.-C. Yuan,
\newblock JCAP {\bf 06}, 030 (2014), 1310.0358.

\bibitem{Ilnicka:2015jba}
A.~Ilnicka, M.~Krawczyk, and T.~Robens,
\newblock Phys. Rev. D {\bf 93}, 055026 (2016), 1508.01671.

\bibitem{Belyaev:2016lok}
A.~Belyaev, G.~Cacciapaglia, I.~P. Ivanov, F.~Rojas-Abatte, and M.~Thomas,
\newblock Phys. Rev. D {\bf 97}, 035011 (2018), 1612.00511.

\bibitem{Kalinowski:2018ylg}
J.~Kalinowski, W.~Kotlarski, T.~Robens, D.~Sokolowska, and A.~F. Zarnecki,
\newblock JHEP {\bf 12}, 081 (2018), 1809.07712.

\bibitem{Engeln:2020fld}
I.~Engeln, P.~Ferreira, M.~M. M\"uhlleitner, R.~Santos, and J.~Wittbrodt,
\newblock JHEP {\bf 08}, 085 (2020), 2004.05382.

\bibitem{Chen:2013jvg}
C.-Y. Chen, M.~Freid, and M.~Sher,
\newblock Phys. Rev. {\bf D89}, 075009 (2014), 1312.3949.

\bibitem{Drozd:2014yla}
A.~Drozd, B.~Grzadkowski, J.~F. Gunion, and Y.~Jiang,
\newblock JHEP {\bf 11}, 105 (2014), 1408.2106.

\bibitem{Muhlleitner:2016mzt}
M.~Mühlleitner, M.~O.~P. Sampaio, R.~Santos, and J.~Wittbrodt,
\newblock JHEP {\bf 03}, 094 (2017), 1612.01309.

\bibitem{Belanger:2021lwd}
G.~Belanger, A.~Mjallal, and A.~Pukhov,
\newblock Phys. Rev. D {\bf 105}, 035018 (2022), 2108.08061.

\bibitem{Belanger:2022qxt}
G.~Belanger, A.~Mjallal, and A.~Pukhov,
\newblock Phys. Rev. D {\bf 106}, 095019 (2022), 2205.04101.

\bibitem{Bhattacharya:2021rwh}
S.~Bhattacharya, S.~Chakraborti, and D.~Pradhan,
\newblock JHEP {\bf 07}, 091 (2022), 2110.06985.

\bibitem{Belanger:2013oya}
G.~Belanger, F.~Boudjema, A.~Pukhov, and A.~Semenov,
\newblock Comput. Phys. Commun. {\bf 185}, 960 (2014), 1305.0237.

\bibitem{Belanger:2018mqt}
G.~Belanger, F.~Boudjema, A.~Goudelis, A.~Pukhov, and B.~Zaldivar,
\newblock Comput. Phys. Commun. {\bf 231}, 173 (2018), 1801.03509.

\bibitem{Bender:2012gc}
C.~M. Bender and S.~Sarkar,
\newblock J. Math. Phys. {\bf 53}, 103509 (2012), 1203.1822.

\bibitem{Edsjo:1997bg}
J.~Edsjo and P.~Gondolo,
\newblock Phys. Rev. D {\bf 56}, 1879 (1997), hep-ph/9704361.

\bibitem{Ma:2006km}
E.~Ma,
\newblock Phys. Rev. {\bf D73}, 077301 (2006), hep-ph/0601225.

\bibitem{Barbieri:2006dq}
R.~Barbieri, L.~J. Hall, and V.~S. Rychkov,
\newblock Phys. Rev. {\bf D74}, 015007 (2006), hep-ph/0603188.

\bibitem{LopezHonorez:2006gr}
L.~Lopez~Honorez, E.~Nezri, J.~F. Oliver, and M.~H.~G. Tytgat,
\newblock JCAP {\bf 0702}, 028 (2007), hep-ph/0612275.

\bibitem{Coimbra:2013qq}
R.~Coimbra, M.~O.~P. Sampaio, and R.~Santos,
\newblock Eur. Phys. J. C {\bf 73}, 2428 (2013), 1301.2599.

\bibitem{Muhlleitner:2020wwk}
M.~M\"uhlleitner, M.~O.~P. Sampaio, R.~Santos, and J.~Wittbrodt,
\newblock Eur. Phys. J. C {\bf 82}, 198 (2022), 2007.02985.

\bibitem{Klimenko:1984qx}
K.~G. Klimenko,
\newblock Theor. Math. Phys. {\bf 62}, 58 (1985).

\bibitem{Hollik:2018wrr}
W.~G. Hollik, G.~Weiglein, and J.~Wittbrodt,
\newblock JHEP {\bf 03}, 109 (2019), 1812.04644.

\bibitem{Ferreira:2019iqb}
{P. M. Ferreira, M. M\"uhlleitner, R. Santos, G. Weiglein, and J. Wittbrodt},
\newblock JHEP {\bf 09}, 006 (2019), 1905.10234.

\bibitem{Evade3}
{Jonas Wittbrodt},
\newblock Evade: Efficient constraints from vacuum decay,
\newblock \url{https://gitlab.com/jonaswittbrodt/EVADE}, 2022,
\newblock [Online; accessed 17-March-2023].

\bibitem{Lee:1977eg}
B.~W. Lee, C.~Quigg, and H.~B. Thacker,
\newblock Phys. Rev. D {\bf 16}, 1519 (1977).

\bibitem{Peskin:1991sw}
M.~E. Peskin and T.~Takeuchi,
\newblock Phys. Rev. D {\bf 46}, 381 (1992).

\bibitem{Grimus:2007if}
W.~Grimus, L.~Lavoura, O.~M. Ogreid, and P.~Osland,
\newblock J. Phys. G {\bf 35}, 075001 (2008), 0711.4022.

\bibitem{Grimus:2008nb}
W.~Grimus, L.~Lavoura, O.~M. Ogreid, and P.~Osland,
\newblock Nucl. Phys. B {\bf 801}, 81 (2008), 0802.4353.

\bibitem{Haller:2018nnx}
J.~Haller {\em et~al.},
\newblock Eur. Phys. J. C {\bf 78}, 675 (2018), 1803.01853.

\bibitem{Bechtle:2013xfa}
P.~Bechtle, S.~Heinemeyer, O.~St\r{a}l, T.~Stefaniak, and G.~Weiglein,
\newblock Eur. Phys. J. C {\bf 74}, 2711 (2014), 1305.1933.

\bibitem{Bechtle:2014ewa}
P.~Bechtle, S.~Heinemeyer, O.~St\r{a}l, T.~Stefaniak, and G.~Weiglein,
\newblock JHEP {\bf 11}, 039 (2014), 1403.1582.

\bibitem{Bechtle:2008jh}
P.~Bechtle, O.~Brein, S.~Heinemeyer, G.~Weiglein, and K.~E. Williams,
\newblock Comput. Phys. Commun. {\bf 181}, 138 (2010), 0811.4169.

\bibitem{Bechtle:2011sb}
P.~Bechtle, O.~Brein, S.~Heinemeyer, G.~Weiglein, and K.~E. Williams,
\newblock Comput. Phys. Commun. {\bf 182}, 2605 (2011), 1102.1898.

\bibitem{Bechtle:2012lvg}
P.~Bechtle {\em et~al.},
\newblock PoS {\bf CHARGED2012}, 024 (2012), 1301.2345.

\bibitem{Bechtle:2015pma}
P.~Bechtle, S.~Heinemeyer, O.~Stal, T.~Stefaniak, and G.~Weiglein,
\newblock Eur. Phys. J. C {\bf 75}, 421 (2015), 1507.06706.

\bibitem{Bahl:2022igd}
H.~Bahl {\em et~al.},
\newblock Comput. Phys. Commun. {\bf 291}, 108803 (2023), 2210.09332.

\bibitem{ATLAS:2022tnm}
ATLAS, G.~Aad {\em et~al.},
\newblock JHEP {\bf 07}, 088 (2023), 2207.00348.

\bibitem{ParticleDataGroup:2022pth}
Particle Data Group, R.~L. Workman {\em et~al.},
\newblock PTEP {\bf 2022}, 083C01 (2022).

\bibitem{OHare:2021utq}
C.~A.~J. O'Hare,
\newblock Phys. Rev. Lett. {\bf 127}, 251802 (2021), 2109.03116.

\bibitem{Azevedo:2018fmj}
D.~Azevedo {\em et~al.},
\newblock JHEP {\bf 11}, 091 (2018), 1807.10322.

\bibitem{Biermann:2022meg}
L.~Biermann, M.~M\"uhlleitner, and J.~M\"uller,
\newblock Eur. Phys. J. C {\bf 83}, 439 (2023), 2204.13425.

\bibitem{Belanger:2020npe}
G.~B\'elanger, C.~Delaunay, A.~Pukhov, and B.~Zaldivar,
\newblock Phys. Rev. D {\bf 102}, 035017 (2020), 2005.06294.

\bibitem{Agrawal:2021dbo}
P.~Agrawal {\em et~al.},
\newblock Eur. Phys. J. C {\bf 81}, 1015 (2021), 2102.12143.

\end{thebibliography}
